\documentclass[aps, prd, onecolumn, tightenlines, notitlepage, superscriptaddress, nofootinbib, preprintnumbers, floatfix,showkeys,11pt,altaffilletter]{revtex4-2}

\usepackage[normalem]{ulem}
\usepackage{amstext}
\usepackage{amssymb}
\usepackage{amsmath}
\usepackage{graphicx}
\graphicspath{{plots/}}
\usepackage{url}
\usepackage{color}
\usepackage{ulem}
\usepackage[utf8]{inputenc}
\pdfoutput=1
\usepackage{textcomp}
\usepackage{comment}
\usepackage{yfonts}
\usepackage{epsfig,amsfonts,mathrsfs,amsmath,amssymb,graphicx,color,slashed,multirow}
\usepackage{amsmath,latexsym,amssymb,graphicx,slashed,color,enumerate,url,cancel,gensymb}
\usepackage{textcomp}

\usepackage[x11names]{xcolor}
\usepackage[colorlinks]{hyperref}

\usepackage{textcase}
\usepackage{amsmath}
\usepackage{booktabs}
\usepackage{adjustbox}

\definecolor{vdrgreen}{rgb}{0.0, 0.7, 0.0}

\definecolor{indianred}{rgb}{0.8, 0.36, 0.36}
\definecolor{blue(ncs)}{rgb}{0.0, 0.53, 0.74}
\AtBeginDocument{\hypersetup{citecolor=indianred,linkcolor=indianred,urlcolor=indianred}}

\usepackage{float}

\usepackage{lmodern}
\usepackage{ae,aecompl}
\usepackage{appendix}
\usepackage{orcidlink}
\usepackage{nccmath} % Provides command \medmath
\usepackage{textcomp}
\usepackage{upgreek}
\usepackage[version=4]{mhchem}
\usepackage[utf8]{inputenc}
\usepackage{fontawesome}
\usepackage{yfonts}
\usepackage[official]{eurosym}
\usepackage{textgreek}


\def\cevns{CE\textnu NS}
\def\d{\mathrm{d}}
\newcommand{\qtransfer}{\left|\mathbf{q}\right|}

\usepackage[T1]{fontenc}
\usepackage{ae,aecompl}
\graphicspath{{Figures/}}
\usepackage{appendix}



\newcommand{\AddrIISERB}{Department of Physics, Indian Institute of Science Education and Research - Bhopal, \\ 
Bhopal Bypass Road, Bhauri, Bhopal 462066, India}

\newcommand{\AddrAHEP}{%
Instituto de F\'{i}sica Corpuscular (IFIC), CSIC‐Universitat de Val\'encia, E-46980 Valencia, Spain}

\usepackage{orcidlink}
\begin{document}

\title{\Large New light mediators and the neutrino fog:\\ Implications from XENONnT nuclear recoil data}

\author{Valentina De Romeri\orcidlink{0000-0003-3585-7437}}\email{deromeri@ific.uv.es}
\affiliation{\AddrAHEP}
\author{Anirban Majumdar~\orcidlink{0000-0002-1229-7951}}\email{anirban19@iiserb.ac.in}
\affiliation{\AddrIISERB}
\author{Dimitrios K. Papoulias~\orcidlink{0000-0003-0453-8492}}\email{dipapou@ific.uv.es}
\affiliation{\AddrAHEP}
\author{Rahul Srivastava~\orcidlink{0000-0001-7023-5727}}\email{rahul@iiserb.ac.in}
\affiliation{\AddrIISERB}


%%%%%%%%%%%%%%%%%%%%%%%%%%%%%%%%%%%%%%%%%%%%%%%%%%%%%
\begin{abstract}

Current ton-scale, xenon-based dark matter (DM) direct detection experiments have now reached the sensitivity required to observe solar neutrinos, marking the onset of the so-called \textit{neutrino fog}. In this work, we explore how this fog is modified when either neutrinos or DM interact with nuclei through a new scalar, vector or axial-vector interaction, considering both heavy and light mediators.
Using the latest nuclear-recoil data from XENONnT, which show indications of coherent elastic neutrino–nucleus scattering from $^8$B solar neutrinos, we derive new strong bounds on couplings of light mediators. We find that these limits are significantly more stringent when the mediator couples to DM, rather than when new physics affects only neutrino interactions. Building on these results, we recompute the expected neutrino fog and compare it with the corresponding constraints on spin-independent and spin-dependent DM-nucleon interactions. We show that the morphology of the neutrino fog can be markedly modified if either neutrinos or DM interact with nuclei through light mediators, even in light of these recent constraints. 

\end{abstract}

\maketitle

%%%%%%%%%%%%%%%%%%%%%%%%%%%%%%%%%%%%%%%%%%%%%%%%%%%%%%%%%%%
\section{\label{Sec:Introduction} Introduction}
%%%%%%%%%%%%%%%%%%%%%%%%%%%%%%%%%%%%%%%%%%%%%%%%%%%%%

Understanding the nature and properties of dark matter (DM)~\cite{Cirelli:2024ssz} remains one of the most pressing challenges in modern physics. Unfortunately, despite extensive experimental and theoretical efforts, no conclusive evidence has yet been found beyond its gravitational effects. Advances in DM direct detection techniques, however, have allowed to reach an impressive precision enabling the testing of very small scattering cross sections~\cite{Schumann:2019eaa,Billard:2021uyg}, under the hypothesis that DM consists of weakly interacting massive particles (WIMPs)~\cite{RevModPhys.90.045002,Cirelli:2024ssz} with masses in the GeV-TeV range. 

The most recent generation of low-threshold dual-phase direct detection experiments, employing liquid xenon  as target material, namely XENONnT~\cite{XENON:2023cxc,XENON:2025vwd}, LZ~\cite{LZ:2024zvo}, and PandaX-4T~\cite{Zhang:2025ajc}, has achieved the highest sensitivities to date, reaching the level required to observe solar neutrinos. Neutrinos can  induce coherent elastic neutrino-nucleus scattering (\cevns)~\cite{Abdullah:2022zue} in the experimental targets, leading to nuclear recoils that can mimic those of DM interactions. In the past few years, the observation of \cevns~at  terrestrial experiments, employing neutrinos from spallation sources (COHERENT~\cite{COHERENT:2017ipa,COHERENT:2020iec, Miranda:2020tif,COHERENT:2021xmm,COHERENT:2025vuz}) or nuclear reactors (Dresden-II~\cite{Colaresi:2022obx} and CONUS+~\cite{Ackermann:2025obx}), has so far validated long-standing theoretical predictions~\cite{Freedman:1973yd,Drukier:1984vhf}. More recently, strong indications of \cevns~ from $^8$B solar neutrinos have also emerged in XENONnT~\cite{XENON:2024ijk}, PandaX-4T~\cite{PandaX:2024muv} and LUX-ZEPLIN (LZ)~\cite{LZ-preprint}.
DM direct detection experiments, marking a milestone for these detectors and posing new challenges for direct DM searches.

Neutrino fluxes originating primarily from astrophysical~\cite{Billard:2013qya}, and to a lesser extent from artificial~\cite{AristizabalSierra:2024smb, Das:2025fle}, sources have long been recognized as an irreducible background for direct DM detection experiments~\cite{Monroe:2007xp,Vergados:2008jp,Strigari:2009bq, Gelmini:2018gqa, Akerib:2022ort}. The recent redefinition of this background in terms of a neutrino fog~\cite{OHare:2021utq} highlights the increasing challenge of distinguishing DM signals from \cevns~events as detector sensitivities continue to improve. At the same time, the observation of \cevns~in DM detectors opens new opportunities to probe the neutrino sector and search for physics beyond the Standard Model (BSM). In this context, recent nuclear recoil data from XENONnT and PandaX-4T have provided a new window into low-energy neutrino physics and have motivated several phenomenological studies exploring possible new interactions and electromagnetic properties of neutrinos~\cite{AristizabalSierra:2024nwf,Li:2024iij,
DeRomeri:2024iaw,DeRomeri:2024hvc,Blanco-Mas:2024ale,Maity:2024aji,Gehrlein:2025isp,AtzoriCorona:2025gyz,Cheek:2025nul}.

Several works in the past have investigated how the expected \cevns~background induced by solar neutrinos in DM direct detection experiments can be modified by the presence of new physics~\cite{Harnik:2012ni,Dent:2016wor,Dent:2016iht,Bertuzzo:2017tuf,Dutta:2017nht,Boehm:2018sux,Schwemberger:2022fjl,Amaral:2023tbs,AristizabalSierra:2017joc,Lozano:2025sug,AristizabalSierra:2021kht}. In this work, we extend previous studies by examining how BSM physics could affect DM direct detection discovery limits in light of recent XENONnT nuclear-recoil data. We explore scenarios featuring new interactions in either the DM or neutrino sectors. We focus on two representative cases. In the first, dubbed case (i), neutrinos interact with nuclei through the SM \cevns~cross section, while DM couples to nuclei via a light scalar, vector or axial-vector mediator. In the second, case (ii), we assume effective spin-independent (SI) or spin-dependent (SD) DM–nucleon interactions (commonly assumed when evaluating the neutrino fog and deriving limits from direct detection data) while allowing neutrinos to have BSM interactions with nuclei mediated by a light scalar, vector or axial-vector mediator. We compare the predicted DM and neutrino event rates with the nuclear recoil data from XENONnT~\cite{XENON:2024ijk}, re-deriving stringent constraints on the SI and SD DM–nucleon cross sections. We then compute the neutrino fog in both scenarios and show how it is modified when accounting for the $^8$B solar neutrino flux measured recently in this experiment and for the BSM interactions under scrutiny. Finally, under the assumption of a light scalar, vector or axial-vector mediator in scenario (i), we obtain strong limits on its coupling to light quarks, which are more restrictive than those derived when the new interaction is assumed to occur in the neutrino sector (scenario (ii), as studied in~\cite{DeRomeri:2024iaw}). 
Notice that similar analyses had been performed in previous works \cite{Dent:2016wor,Boehm:2018sux,AristizabalSierra:2021kht,Bertuzzo:2017tuf}, although following a neutrino-floor approach. Here, we update and extend these results in two ways: first, we use the  neutrino fog approach to evaluate discovery limits, and second, we account for the most recent bounds on the new mediators, from the first XENONnT \cevns~data.

Our paper is organized as follows. We provide in Sec.~\ref{Sec:Theory} the theoretical framework to compute the DM and neutrino event rates in the two scenarios. Section~\ref{Sec:Data_Analysis} is devoted to the discussion of the statistical analysis method and comparison with experimental data. In Sec.~\ref{Sec:Neutrino_Fog} we discuss how we compute the neutrino fog. We present our results in terms of bounds on the DM-nuclei cross sections, new mediator couplings and the recomputed neutrino fog in Sec.~\ref{Sec:Results}. Finally, we draw our conclusion in Sec.~\ref{Sec:Conclusions}.

%%%%%%%%%%%%%%%%%%%%%%%%%%%%%%%%%%%%%%%%%%%%%%%%%%%%%%%%%%%%%%%%%%%%%%%%%%%%%%%%
\section{\label{Sec:Theory}Theoretical Framework}
%%%%%%%%%%%%%%%%%%%%%%%%%%%%%%%%%%%%%%%%%%%%%%%%%%%%%%%%%%%%%%%%%%%%%%%%%%%%%%%%

In this section, we outline the theoretical framework used to compute  DM and neutrino event rates in DM direct detection experiments for the two BSM scenarios under consideration.

%%%%%%%%%%%%%%%%%%%%%%%%%%%%%%%%%%%%%%%%%%%%%%%%%%%%%%%%%%%%%%%%%%%%%%%%%%%%%%%%
\subsection{\label{SubSec:DM_N_Interactions}Dark matter–nucleus scattering}
%%%%%%%%%%%%%%%%%%%%%%%%%%%%%%%%%%%%%%%%%%%%%%%%%%%%%%%%%%%%%%%%%%%%%%%%%%%%%%%%

We start by defining the DM-nucleus interaction. The differential rate of nuclear recoil events for DM–nuclei scattering is given by~\cite{Lewin:1995rx, Freese:2012xd}

\begin{equation}
\label{Eq.DM_Nuclear_Scattering}
\left[\frac{\d R}{\d T_\mathcal{N}}\right]^X_{\chi \mathcal{N}} = t_\mathrm{run} N_\mathrm{target} n_\chi \int_{v_\mathrm{min}}^{v_\mathrm{max}} \d^3v \, v f(\mathbf{v}) \frac{\d\sigma^X_{\chi \mathcal{N}}}{\d T_\mathcal{N}}\,,
\end{equation}
where $\chi$ denotes a generic  spin-1/2 WIMP candidate and $\mathcal{N}$ the target nucleus; moreover, $t_\mathrm{run}$ is the experiment exposure time and $N_\mathrm{target} = m_\mathrm{det} N_A / m_\mathrm{target}$ is the number of target nuclei in the detector, while $X$ denotes the nature of the BSM interaction.  Here, $N_A$ denotes Avogadro’s number, $m_\mathrm{target}$ is the molar mass of the target nuclei, and $m_\mathrm{det}$ is the detector fiducial mass. The rate accounts also for the WIMP number density $n_\chi = \rho_\chi / m_\chi$, where we adopt the Standard Halo Model (SHM) with $\rho_\chi =  0.3~\mathrm{GeV/cm}^3$ fixed to the local DM density and $m_\chi$ the WIMP mass. The lower limit of the velocity integral, $v_\mathrm{min}$, is set by the minimum DM speed required to transfer a nuclear recoil energy $T_\mathcal{N}$ in an elastic, non-relativistic DM-nuclei scattering. From kinematics, the minimum DM speed reads
\begin{equation}
v_\mathrm{min} = \sqrt{\frac{m_\mathcal{N} T_\mathcal{N}}{2\,\mu_{\chi\mathcal{N}}^{\,2}}}\,,
\end{equation}
with $\mu_{\chi\mathcal{N}} = \frac{m_\chi m_\mathcal{N}}{m_\chi + m_\mathcal{N}}$ being the $\chi$–$\mathcal{N}$ reduced mass and $m_\mathcal{N}$ the nucleus mass. 

The WIMP velocity distribution $f(\mathbf{v})$ in the laboratory frame is obtained via the Galilean transformation $f(\mathbf{v}) = \tilde{f}(\mathbf{v}+\mathbf{v}_\mathrm{lab})$, where $\tilde{f}(\mathbf{v})$ is the distribution in the DM rest frame and $\mathbf{v}_\mathrm{lab} = 232~\mathrm{km/s}$ is the lab velocity relative to the local DM rest frame. Under the assumption of SHM, the local DM halo is assumed to be smooth, virialized, and isotropic, with a Maxwellian velocity distribution of root mean square (rms) dispersion $\sigma_v$. The distribution in the DM rest frame is

\begin{equation}
\tilde{f}(\mathbf{v}) = \frac{1}{N_\mathrm{esc}} \left( \frac{3}{2\pi\sigma_v^2} \right)^{3/2} e^{-3v^2 / (2\sigma_v^2)} \Theta(v_\mathrm{esc} - |\mathbf{v}|)\,,
\end{equation}
where the Heaviside function $\Theta(x)$ enforces the cutoff at the Galactic escape velocity $v_\mathrm{esc} = 544~\mathrm{km/s}$. The normalization factor $N_\mathrm{esc}$ is

\begin{equation}
N_\mathrm{esc} = \mathrm{erf}\left[\frac{v_\mathrm{esc}}{v_{_0}}\right] - \frac{2}{\sqrt{\pi}}\frac{v_\mathrm{esc}}{v_{_0}} e^{-(v_\mathrm{esc}/v_{_0})^2}\,,
\end{equation}
where the local circular speed $v_{_0} = 220~\mathrm{km/s}$ is correlated with $\sigma_v$ through $v_{_0} = \sqrt{2/3} \, \sigma_v$.\\

The differential cross section $\d\sigma^X_{\chi \mathcal{N}}/\d T_\mathcal{N}$ encodes the particle physics details. We consider three cases of Lorentz-invariant bilinear effective interactions for spin-$1/2$ DM, namely scalar ($S$), vector ($V$), and axial-vector ($A$) mediators, leading to elastic $\chi$–$\mathcal{N}$ scattering via tree-level exchange. The generic effective Lagrangian for DM–quark interactions is

\begin{equation}
\mathscr{L}_{\chi q}^X \subset g_\chi^X \overline{\chi} \Gamma^X \chi \, X + g_q^X \overline{q} \Gamma_X q \, X\,,
\end{equation}
where $\Gamma^X = \{\mathbb{I}, \gamma^\mu, \gamma^\mu \gamma_5\}$ correspond to $X = S, V, A$, respectively, and $g_{\chi}^X$, $g_{q}^X$ are dimensionless couplings. Scalar and vector mediators produce SI interactions, while the axial yields SD interactions~\cite{DelNobile:2021wmp,Fitzpatrick:2012ib,Fitzpatrick:2012ix}.

To obtain the DM–nucleus scattering cross sections from the quark-level interactions, we follow Ref.~\cite{DelNobile:2021wmp}. In the non-relativistic limit, the $\chi$–$\mathcal{N}$ differential cross sections are

\begin{align}
\label{eq:DM_nuclei_xsec_newphysicsS}
\frac{\d\sigma_{\chi \mathcal{N}}^S}{\d T_\mathcal{N}} &= \frac{m_\mathcal{N}}{2\pi v^2} \cdot \frac{\left(Q_{\chi\mathcal{N}}^S\right)^2}{(\qtransfer^2 + m_S^2)^2}\,,\\
\label{eq:DM_nuclei_xsec_newphysicsV}
\frac{\d\sigma_{\chi \mathcal{N}}^V}{\d T_\mathcal{N}} &= \frac{m_\mathcal{N}}{2\pi v^2} \cdot \frac{\left(Q_{\chi\mathcal{N}}^V\right)^2}{(\qtransfer^2 + m_V^2)^2}\,,\\
\label{eq:DM_nuclei_xsec_newphysicsA}
\frac{\d\sigma_{\chi \mathcal{N}}^A}{\d T_\mathcal{N}} &= \frac{8m_\mathcal{N}}{v^2} \cdot \frac{1}{2\mathcal{J}+1} \cdot \frac{\left(Q_{\chi\mathcal{N}}^A\right)^2}{(\qtransfer^2 + m_A^2)^2}\,,
\end{align}
with $\qtransfer = \sqrt{2 m_\mathcal{N} T_\mathcal{N}}$ denoting the momentum transfer and $\mathcal{J}$  the nuclear spin.
The effective charges $Q_{\chi\mathcal{N}}^X$ for $X = S, V, A$ are~\cite{DelNobile:2021wmp}

\begin{align}
Q_{\chi\mathcal{N}}^S &= g_\chi^S \left[ Z \sum_q g_q^S \frac{m_p}{m_q} f_{Tq}^p + N \sum_q g_q^S \frac{m_n}{m_q} f_{Tq}^n \right] F_\mathrm{SI}(\qtransfer^2)\,,\\
Q_{\chi\mathcal{N}}^V &= g_\chi^V \left[ Z(2g_u^V + g_d^V) + N(g_u^V + 2g_d^V) \right] F_\mathrm{SI}(\qtransfer^2)\,,\\
\left(Q_{\chi\mathcal{N}}^A\right)^2 &= (g_{\chi}^A)^2 \left[(g_{0}^A)^2 \mathcal{S}_{00}(\qtransfer^2)+(g_{0}^A)^2 \mathcal{S}_{11}(\qtransfer^2)+g_{1}^Ag_1^A \mathcal{S}_{01}(\qtransfer^2)\right]\,,\label{Eq:Axial_Charge_DM}
\end{align}
where $Z$ and $N$ stand for the number of protons and neutrons, $m_p$ and $m_n$ are the proton and neutron masses, and $m_q$ are the light quark masses ($q = \{u, d, s\}$).  The scalar hadronic matrix elements are fixed to~\cite{DelNobile:2021wmp, FlavourLatticeAveragingGroup:2019iem}
\begin{align*}
f_{Tu}^p \approx 0.026; \quad f_{Td}^p \approx 0.038; \quad f_{Ts}^p \approx 0.044;\\
f_{Tu}^n \approx 0.018; \quad f_{Td}^n \approx 0.056; \quad f_{Ts}^n \approx 0.044.
\end{align*}
For axial-vector interactions, the isoscalar and isovector couplings are defined as $g_0^A = (g_p^A+g_n^A)/2$ and $g_1^A = (g_p^A-g_n^A)/2$, respectively, with $g_p^A = \sum_q g_q^A \Delta_q^p$ and $g_n^A = \sum_q g_q^A \Delta_q^n$. Here, $\Delta_{q}^\mathfrak{n}$ parametrize the quark spin content of the nucleon and under the assumption of isospin symmetry the values are fixed from lattice QCD studies~\cite{Lin:2018obj} as
\begin{align*}
\Delta_u^p = \Delta_d^n \approx 0.777; &&&&&&& \Delta_d^p = \Delta_u^n \approx -0.438; &&&&&&& \Delta_s^p = \Delta_s^n \approx -0.053.
\end{align*}

Finite nuclear size effects are included via nuclear form factors. Specifically, for the case of $S$ and $V$ interactions, the Helm parametrization~\cite{Helm:1956zz} is adopted

\begin{equation}
\label{Eq.Helm}
F_\mathrm{SI}(\qtransfer^2) = \frac{3 j_1(\qtransfer R_0)}{\qtransfer R_0} e^{-\frac{1}{2} (\qtransfer s)^2}\,,
\end{equation}
where $j_1(x)$ is the spherical Bessel function of order one, with
\begin{align*}
R_0 = \sqrt{\frac{5}{3} \langle R^2 \rangle - 5s^2}, \quad \langle R^2 \rangle = 3/5 R^2, \quad R = 1.23 A^{1/3}~\mathrm{fm}, \quad s = 0.9~\mathrm{fm}.
\end{align*}

On the other hand, for the case of axial-vector interactions, Eq.~\eqref{Eq:Axial_Charge_DM} contains the nuclear spin structure functions $\mathcal{S}_{ij}(\qtransfer^2)$ $(i,j=0,1)$. These can be evaluated by adding the transverse, $\mathcal{S}^\mathcal{T}_{ij}(\qtransfer^2)$, and longitudinal, $\mathcal{S}^\mathcal{L}_{ij}(\qtransfer^2)$ components, following the prescriptions derived in Ref.~\cite{Hoferichter:2020osn}. In particular, we adopt the nuclear spin structure functions and their momentum-transfer dependence as given in Sec.~E and Table~VIII of Ref.~\cite{Hoferichter:2020osn} (see also Fig.~8 therein for the corresponding spin structure functions of $^{129}$Xe and $^{131}$Xe, which illustrate their $\qtransfer$ dependence). For the nuclear target ground-state to ground-state interactions considered in the present work, the SD contributions arise only from the odd-$A$ isotopes $^{129}$Xe and $^{131}$Xe, for which $\mathcal{J}\neq0$.

Before closing this discussion, let us note that direct detection constraints are often quoted in terms of DM–nucleon cross sections at the zero-momentum transfer limit, $\qtransfer^2 \to 0$. The total $\chi$–nucleon cross sections in this limit are~\cite{DelNobile:2021wmp}

\begin{align}
\label{eq:lightmedlimit_DMxsecs}
\left.\sigma_{\chi \mathfrak{n}}^\mathrm{SI}\right|_{\qtransfer^2 \to 0} &= \frac{\mu^2_{\chi \mathfrak{n}}}{\pi} \frac{\left( g_\chi^{V/S} g_\mathfrak{n}^{V/S} \right)^2}{m_{V/S}^4},\\
\left.\sigma_{\chi \mathfrak{n}}^\mathrm{SD}\right|_{\qtransfer^2 \to 0} &= \frac{3 \mu^2_{\chi \mathfrak{n}}}{\pi} \frac{\left( g_\chi^A g_\mathfrak{n}^A \right)^2}{m_A^4},
\end{align}
where $\mu_{\chi \mathfrak{n}} = m_\chi m_\mathfrak{n} / (m_\chi + m_\mathfrak{n})$ is the DM–nucleon reduced mass ($m_\mathfrak{n}$ is either $m_p$ or $m_n$), and the nucleon couplings are $g_\mathfrak{n}^S = \sum_q ~g_q^S~m_\mathfrak{n}/m_q ~f_{Tq}^\mathfrak{n}$, $g_\mathfrak{n}^V = \sum_q g_q^V~\mathscr{N}_q^\mathfrak{n}$ and $g_\mathfrak{n}^A = \sum_q g_q^A~\Delta_q^\mathfrak{n}$, with $\mathscr{N}_q^\mathfrak{n}$ being the number of valence quarks in the nucleon. Notably, in the SI case, rewriting Eqs.~\eqref{eq:DM_nuclei_xsec_newphysicsS} and \eqref{eq:DM_nuclei_xsec_newphysicsV} in terms of the zero–momentum DM–nucleon cross section shows that scalar and vector mediators yield the same analytic form for $\mathrm{d}\sigma_{\chi\mathcal N}/\mathrm{d}T_{\mathcal N}$, exhibiting identical functional dependence on the parameters $(m_\chi, m_X, \sigma_{\chi\mathfrak n}^{\mathrm{SI}}|_{\qtransfer^2\to0})$; the differences in the underlying hadronic currents are fully absorbed into $\sigma_{\chi\mathfrak n}^{\mathrm{SI}}|_{\qtransfer^2\to0}$. For completeness, we note that other Lorentz structures, namely pseudoscalar and tensor interactions, can, in principle, contribute. However, pseudoscalar interactions are strongly momentum-suppressed in DM direct detection, while tensor interactions effectively generate SD responses similar to the axial-vector case, differing only by a numerical factor of $\sim 4$ at the cross-section level, neglecting a sub-leading contribution proportional to the component of the nucleon spin longitudinal to $\qtransfer$~\cite{DelNobile:2021wmp,Fitzpatrick:2012ix}.

%%%%%%%%%%%%%%%%%%%%%%%%%%%%%%%%%%%%%%%%%%%%%%%%%%%%%%%%%%%%%%%%%%%%%%%%%%%%%%%%
\subsection{\label{SubSec:CEvNS_Interactions}
Coherent elastic neutrino-nucleus scattering}
%%%%%%%%%%%%%%%%%%%%%%%%%%%%%%%%%%%%%%%%%%%%%%%%%%%%%%%%%%%%%%%%%%%%%%%%%%%%%%%%

We now turn our attention to another significant aspect of direct detection experiments: the neutrino background. Astrophysical neutrinos constitute the dominant background in direct detection searches, representing an irreducible component that is inherently linked to the observations. The primary channel of interaction for neutrinos in the context of DM direct searches through nuclear recoils is \cevns.
Neutrino backgrounds relevant to DM direct detection searches are typically categorized into contributions from solar neutrinos~\cite{Haxton:2012wfz}, atmospheric neutrinos~\cite{Battistoni:2005pd}, and the diffuse supernova neutrino background (DSNB)~\cite{Yuksel:2008cu, Beacom:2010kk}. Geo-\cite{2013arXiv1301.0365H, Gelmini:2018gqa,Kosmas:2021zve} and reactor neutrinos~\cite{Baldoncini:2014vda,AristizabalSierra:2024smb,Das:2025fle} may also contribute to this irreducible background, albeit at a subdominant level. Note that given the current energy thresholds and sensitivities of leading direct detection experiments, only $^8$B solar neutrinos are expected to produce nuclear recoil energies within the detectable range.
 
Recent indications of $^{8}$B solar neutrino–induced \cevns~events reported by the XENONnT and PandaX-4T collaborations have provided the first experimental confirmation of these expectations. These results highlight the importance of accurately characterizing this background in order to ensure a robust interpretation of experimental data, both for a possible DM detection and to establish solid exclusion limits. These observations yield $^{8}$B flux normalizations of $(4.7^{+3.6}_{-2.3}) \times 10^{6}~\mathrm{cm^{-2} s^{-1}}$~\cite{XENON:2024ijk} and $(8.4 \pm 3.1) \times 10^{6}~\mathrm{cm^{-2} s^{-1}}$~\cite{PandaX:2024muv}, respectively. Although not yet competitive with dedicated solar neutrino measurements~\cite{SNO:2011hxd,Borexino:2017uhp,SNO:2018fch,Super-Kamiokande:2023jbt,ParticleDataGroup:2024cfk}, these results provide an increasingly precise characterization of the dominant neutrino background relevant for DM masses in the few-GeV range, and represent an important step toward  characterizing the neutrino fog with current-generation detectors.

\cevns~is a neutral-current  interaction in which a neutrino scatters coherently off an entire nucleus. The differential nuclear recoil rate corresponding to \cevns~from a neutrino source $j$ reads~\cite{Papoulias:2018uzy}

\begin{equation}
\label{Eq.cevns_diff_events}
\left[\frac{\d R}{\d T_\mathcal{N}}\right]^X_{\nu \mathcal{N}}
= t_\mathrm{run} \, N_\mathrm{target}
\int_{E_\nu^\mathrm{min}}^{E_\nu^\mathrm{max}} \d E_\nu \,
\frac{\d\Phi^{j}_\nu(E_\nu)}{\d E_\nu} \,
\frac{\d\sigma^X_{\nu \mathcal{N}}}{\d T_\mathcal{N}} \,,
\end{equation}
where $\frac{\d\Phi^{j}_\nu(E_\nu)}{\d E_\nu}$ is the energy spectrum of the neutrino flux $j$, and $\frac{\d\sigma^X_{\nu \mathcal{N}}}{\d T_\mathcal{N}} $ denotes the differential \cevns\ cross section with $X=\mathrm{SM}$ for the SM-only case, or $X=\mathrm{SM}$ plus the contribution of a new $S$, $V$ or $A$ mediator. The integration limits $E_\nu^\mathrm{min}$ ($E_\nu^\mathrm{max}=16.36 \mathrm{~MeV}$, in the case of $j = ^{8}$B solar neutrinos) denotes the minimum (maximum) neutrino energy of the  neutrino source, respectively. From kinematic considerations, the minimum neutrino energy capable of producing a nuclear recoil of energy $T_\mathcal{N}$ is given by

\begin{equation}
E_\nu^\mathrm{min}=\frac{1}{2}\left[T_\mathcal{N}+\sqrt{T_\mathcal{N}^2+2m_{\mathcal{N}}T_\mathcal{N}}\right]\,.
\end{equation}

Within the SM, the differential \cevns~cross section can be expressed as~\cite{Freedman:1977xn}

\begin{equation}
\label{Eq.CEvNS_SM_xSec}
\frac{\d\sigma^\mathrm{SM}_{\nu \mathcal{N}}}{\d T_\mathcal{N}}=\frac{G_F^2 m_\mathcal{N}}{\pi}{\left(Q_{\nu \mathcal{N}}^\mathrm{SM}\right)}^2\left(1-\frac{m_\mathcal{N}T_\mathcal{N}}{2 E_\nu^2}\right)\,,
\end{equation}
where $G_F$ is the Fermi constant, while $Q_{\nu \mathcal{N}}^\mathrm{SM}$ is the SM coherent weak charge. The latter quantity can be written in terms of the SM weak neutrino–quark couplings as

\begin{equation} 
\label{Eq.CEvNS_Coupling} 
Q_{\nu \mathcal{N}}^\mathrm{SM} = [N(g^\mathrm{SM}_u+2g^\mathrm{SM}_d)+Z(2g^\mathrm{SM}_u+g^\mathrm{SM}_d)]F(\qtransfer^2)\,, 
\end{equation} %
where the tree-level couplings are $g_u^\mathrm{SM} = \tfrac{1}{2} - \tfrac{4}{3}\sin^2\theta_W$ and $g_d^\mathrm{SM} = -\tfrac{1}{2} + \tfrac{2}{3}\sin^2\theta_W$ \footnote{Radiative corrections introduce small modifications~\cite{AtzoriCorona:2024rtv,Erler:2013xha}, requiring huge exposures to become relevant~\cite{Mishra:2023jlq}.}, and where $\sin^2\theta_W = 0.2387$~\cite{ParticleDataGroup:2024cfk} is the weak mixing angle. The nuclear form factor $F(\qtransfer^2)$ is taken to be identical for neutrons and protons and is modeled using the Helm parametrization of Eq.~\eqref{Eq.Helm}, an excellent approximation given the small momentum transfers relevant for \cevns\ induced by solar $^8$B neutrinos.\\

In the presence of BSM interactions the \cevns~cross section will be modified, with the addition of new terms that may interfere with SM one, or add incoherently, depending on the nature of the new interaction.
We consider the following effective Lagrangian\footnote{In Eq.~\eqref{Eq.BSM_CEvNS_Lagrangian} the neutrino field $\nu$ is treated as a four–component Dirac spinor, $\nu = \nu_L + \nu_R$. Depending on the Lorentz structure $\Gamma^X$, only the chirality–allowed bilinears contribute. For example, for a scalar mediator ($\Gamma^S = \mathbb{I}$) the coupling connects opposite chiralities, effectively $\overline{\nu}\nu\sim\overline{\nu}_L\nu_R$, whereas for a vector ($\Gamma^V = \gamma^\mu$) or an axial-vector  ($\Gamma^{A} = \gamma^\mu \gamma_5$) mediator, only the left–handed current $\overline{\nu}_L\gamma^\mu \nu_L$ and $\overline{\nu}_L\gamma^\mu \gamma_5\nu_L$  contributes, respectively.} coupling new mediators to quarks, $q$, and neutrinos, $\nu$,

\begin{equation}
\label{Eq.BSM_CEvNS_Lagrangian}
\mathscr{L}_{\nu q}^X \subset g_{\nu}^X\overline{\nu}\Gamma^X\nu X+ g_{q}^X\overline{q}\Gamma^X q X\,,
\end{equation}
where $\Gamma^X = \{\mathbb{I}, \gamma^\mu, \gamma^\mu \gamma_5\}$, $X$ corresponding to $S, V, A$, respectively, while $g_{\nu}^X$ and $g_{q}^X$ represent dimensionless coupling constants of the new mediator with neutrinos and the first generation of quarks, respectively. 
The differential \cevns~cross sections in the presence of SM interactions plus an additional scalar, vector, or axial-vector mediator can be written as~\cite{Cerdeno:2016sfi, AristizabalSierra:2018eqm,Candela:2024ljb}

\begin{align}
\label{Eq.SM_plus_S_CEvNS_xSec}
\frac{\d\sigma^S_{\nu \mathcal{N}}}{\d T_\mathcal{N}} &= \frac{\d\sigma^\mathrm{SM}_{\nu \mathcal{N}}}{\d T_\mathcal{N}}
+ \left(Q_{\nu\mathcal{N}}^S\right)^2
\frac{m_\mathcal{N}^2 T_\mathcal{N}}{4\pi E_\nu^2 \left(\qtransfer^2 + m_S^2\right)^2}\,,\\[4pt]
%%%%%%%%%%%%%
\label{Eq.SM_plus_V_CEvNS_xSec}
\frac{\d\sigma^V_{\nu \mathcal{N}}}{\d T_\mathcal{N}} &= \frac{G_F^2 m_\mathcal{N}}{\pi}
\left[ Q_{\nu\mathcal{N}}^\mathrm{SM} + \kappa \frac{Q_{\nu\mathcal{N}}^V}{\sqrt{2} G_F \left(\qtransfer^2 + m_V^2\right)} \right]^2
\left( 1 - \frac{m_\mathcal{N} T_\mathcal{N}}{2E_\nu^2} \right)\,,\\[4pt]
%%%%%%%%%%%%%%%%%%
\label{Eq.SM_plus_A_CEvNS_xSec}
\frac{\d\sigma^A_{\nu \mathcal{N}}}{\d T_\mathcal{N}} &= \frac{\d\sigma^\mathrm{SM}_{\nu \mathcal{N}}}{\d T_\mathcal{N}}
+ \left(Q_{\nu\mathcal{N}}^A\right)^2
\frac{4m_\mathcal{N}}{(2\mathcal{J}+1)\left(\qtransfer^2 + m_A^2\right)^2}
\left( 1 + \frac{m_\mathcal{N} T_\mathcal{N}}{2E_\nu^2} \right)\,.
\end{align}
The corresponding effective charges $Q_{\nu\mathcal{N}}^X$ $(X = S, V, A)$ are given by

\begin{align}
Q_{\nu\mathcal{N}}^S &= g_\nu^S \left[ Z \sum_q g_q^S \frac{m_p}{m_q} f_{Tq}^p + N \sum_q g_q^S \frac{m_n}{m_q} f_{Tq}^n \right] F(\qtransfer^2)\,,\\
\label{Eq.vector_BSM_charge}
Q_{\nu\mathcal{N}}^V &= g_\nu^V \left[ Z(2g_u^V + g_d^V) + N(g_u^V + 2g_d^V) \right] F(\qtransfer^2)\,,\\
\left(Q_{\nu\mathcal{N}}^A\right)^2 &= \left(g_\nu^A\right)^2 \left[ \left(g_0^A\right)^2 \mathcal{S}_{00}^\mathcal{T}(\qtransfer^2) + g_0^A g_1^A \mathcal{S}_{01}^\mathcal{T}(\qtransfer^2) + \left(g_1^A\right)^2 \mathcal{S}_{11}^\mathcal{T}(\qtransfer^2) \right]\,.
\end{align}

For axial-vector interactions, the isoscalar and isovector couplings are defined as $g_0^A = (g_p^A+g_n^A)/2$ and $g_1^A = (g_p^A-g_n^A)/2$, with $g_p^A = \sum_q g_q^A \Delta_q^p$ and $g_n^A = \sum_q g_q^A \Delta_q^n$. The nuclear spin structure functions $\mathcal{S}_{ij}^\mathcal{T}(\qtransfer^2)$ $(i,j=0,1)$ are the transverse spin structure functions\footnote{The terms proportional to the longitudinal component of the axial-vector cross section  scale as $T_\mathcal{N}/m_\mathcal{N}$, hence they are very suppressed and can be safely ignored~\cite{Hoferichter:2020osn}.}, evaluated following the prescriptions from Ref.~\cite{Hoferichter:2020osn} (see also Refs.~\cite{Candela:2024ljb,DeRomeri:2024iaw, Chattaraj:2025rtj}). Notice that the axial interaction is suppressed due to its spin-dependent nature; hence, it does not lead to the typical coherent enhancement found for the vector case. Moreover, in a xenon target, only the odd-neutron isotopes $^{129}$Xe and $^{131}$Xe have non-zero nuclear spin and thus contribute to the axial channel. Their combined natural abundance of about 50\% and the absence of coherent enhancement significantly suppress the spin-dependent signal.

Finally, for vector interactions, we consider two possibilities in Eq.~\eqref{Eq.SM_plus_V_CEvNS_xSec}: a mediator with universal couplings to neutrinos and quarks, in which case $\kappa = 1$. This is a commonly used phenomenological scenario used to provide an indicative comparison of experimental sensitivities, which, however, is not anomaly-free. For this reason, we also consider a 
$U(1)_{B\!-\!L}$ extension of the SM, for which the corresponding weak charge has $\kappa = -1/3$.
Note that given the conventions used in our definition, $Q_{\nu \mathcal{N}}^\mathrm{SM}$ is always negative. The $B\!-\!L$ weak charge is also always negative and therefore cannot lead to 
destructive interference with the SM contribution, in contrast to the 
universal mediator case. As will become evident in the results shown below, 
this distinction has clear phenomenological implications.

%%%%%%%%%%%%%%%%%%%%%%%%%%%%%%%%%%%%%%%%%%%%%%%%%%%%%%%%%%%%%%%%%%%%%%%%%%%%%%%%
\section{\label{Sec:Data_Analysis} Simulation of event spectra and statistical analysis}
%%%%%%%%%%%%%%%%%%%%%%%%%%%%%%%%%%%%%%%%%%%%%%%%%%%%%%%%%%%%%%%%%%%%%%%%%%%%%%%%

In this section, we provide a brief overview of the statistical framework adopted to analyze the recent XENONnT data~\cite{XENON:2024ijk}. Although not discussed here, we note that performing a similar analysis using PandaX-4T data~\cite{PandaX:2024muv,Zhang:2025ajc} would lead to analogous conclusions\footnote{This was also observed in previous \cevns-oriented studies, which found comparable constraints in various BSM scenarios~\cite{DeRomeri:2024hvc,DeRomeri:2024iaw,Blanco-Mas:2024ale,AtzoriCorona:2025gyz}.}. Indeed, the XENONnT and PandaX-4T experiments have collected data on xenon targets with nearly identical exposures, thresholds, and regions of interest. More importantly, both experiments have measured the $^8$B solar neutrino flux with sizable statistical uncertainties, 63\% for XENONnT and 37\% for PandaX-4T, albeit with different best-fit normalizations.
For the purposes of investigating the neutrino fog, the dominant factor shaping it is the size of the $^8$B flux uncertainty rather than its absolute normalization. We therefore base our analysis on the XENONnT dataset as a more conservative choice, further noting that its $^8$B flux normalization is closer to those measured by dedicated solar neutrino experiments such as SNO and Borexino.

Special emphasis is devoted to the aspects relevant for simulating the expected event spectra from both DM-nucleus and \cevns~interactions in two-phase liquid xenon time projection (TPC LXe) chambers. In such detectors, an energy deposition in the LXe produces two measurable signals: prompt scintillation photons (S1) and delayed electroluminescence photons (S2) generated by ionization electrons extracted into the gaseous phase. The combined analysis of these signals enables effective discrimination between nuclear and electron recoil events.

In Sec.~\ref{Sec:Theory}, we outlined the procedure to compute the theoretical differential event rates for DM-nucleus scattering and \cevns, given in Eqs.~\eqref{Eq.DM_Nuclear_Scattering} and \eqref{Eq.cevns_diff_events}, respectively. In the following subsections, we describe how these theoretical rates are translated into realistic signal predictions for the XENONnT experiment, taking into account detector-specific efficiencies and other experimental response effects.

%%%%%%%%%%%%%%%%%%%%%%%%%%%%%%%%%%%%%%%%%%%%%%%%%%%%%%%%%%%%%%%%%%%%%%%%%%%%%%%%
\subsection{\label{SubSec:XENONnT_Analysis}Event simulation and statistical analysis for XENONnT}
\label{subsec: XENONnT_analysis}
%%%%%%%%%%%%%%%%%%%%%%%%%%%%%%%%%%%%%%%%%%%%%%%%%%%%%%%%%%%%%%%%%%%%%%%%%%%%%%%%

We consider the reported dataset~\cite{XENON:2024ijk} obtained with an exposure of $3.51~\text{tonne year}$, consistent with the most recent dataset reported by the collaboration. Data are presented in three (corrected) S2 bins within the range $[120, 500]$ photoelectrons (PE). The expected number of DM $(\chi \mathcal{N}$) or \cevns~$(\nu \mathcal{N}$) nuclear recoil events in the $i$-th bin is obtained as

\begin{equation}
\label{Eq:XENONnT_Events}
R_{\nu \mathcal{N} (\chi \mathcal{N})}^{i,X} = c^i \int_{\mathrm{S2}^i}^{\mathrm{S2}^{i+1}} \d\mathrm{S2}\,
\mathcal{A}^\mathrm{XENONnT}(T_\mathcal{N}) \,
\left [\frac{\d R}{\d T_\mathcal{N}} \right]_{\nu \mathcal{N} (\chi \mathcal{N})}^X \,
\frac{\d T_\mathcal{N}}{\d\mathrm{S2}}\,,
\end{equation}
where $\mathcal{A}^\mathrm{XENONnT}(T_\mathcal{N})$ is the experimental detection efficiency extracted from Refs.~\cite{XENON:2024ijk,XENON:2024hup}, for the combined S1 + S2 datasets. Following Ref.~\cite{DeRomeri:2024iaw}, we include the correction factors $c^i$ in Eq.~\eqref{Eq:XENONnT_Events}, having the role of effective efficiencies, in order to align our predictions with the best-fit spectra reported by the XENONnT collaboration (see the first panel of Fig.~2 in Ref.~\cite{XENON:2024ijk}). The mapping between the nuclear recoil energy $T_\mathcal{N}$ and the S2 observable is performed through

\begin{equation}
\mathrm{S2} = T_\mathcal{N} \, Q_y(T_\mathcal{N}) \, g_2\,,
\end{equation}
with $g_2 = 16.9~\mathrm{PE/e^{-}}$ denoting the single-electron amplification gain~\cite{XENON:2024ijk} and $Q_y(T_\mathcal{N})$ the charge yield function, taken from~Ref.~\cite{XENON:2024kbh}.  

The statistical analysis of the XENONnT data is carried out via a binned spectral analysis, based on the Poisson-likelihood $\chi^2$ function

\begin{align}
\chi^2(\overrightarrow{\mathcal{S}};\alpha,\beta,\gamma_k) &= 2 \sum_i \Biggl[ 
R_\mathrm{pred}^i(\overrightarrow{\mathcal{S}};\alpha,\beta,\gamma_k) - R_\mathrm{exp}^i 
+ R_\mathrm{exp}^i \ln \left( \frac{R_\mathrm{exp}^i}{R_\mathrm{pred}^i(\overrightarrow{\mathcal{S}};\alpha,\beta,\gamma_k)} \right)
\Biggr] \notag \\
&\quad + \left( \frac{\alpha}{\sigma_\alpha} \right)^2 
+ \left( \frac{\beta}{\sigma_\beta} \right)^2
+ \sum_k \left( \frac{\gamma_k}{\sigma_{\gamma_k}} \right)^2\,.
\end{align}
Here, $R_\mathrm{exp}^i$ denotes the observed number of events in the $i$-th bin (from Ref.~\cite{XENON:2024ijk}), while the predicted number of events is obtained as

\begin{align}
R_\mathrm{pred}^i(\overrightarrow{\mathcal{S}};\alpha,\beta,\gamma_k) &= (1+\alpha)\, R_{\chi\mathcal{N}}^i(\overrightarrow{\mathcal{S}}) 
+ (1+\alpha+\beta)\, R_{\nu \mathcal{N}}^i(\overrightarrow{\mathcal{S}}) \notag \\
&\quad + (1+\gamma_1)\, R_\mathrm{AC}^i 
+ (1+\gamma_2)\, R_\mathrm{neutron}^i
+ (1+\gamma_3)\, R_\mathrm{ER}^i \,.
\end{align}
The nuisance parameter $\alpha$ encodes a $5\%$ uncertainty on the fiducial volume of the detector, while $\beta$ accounts for the $4\%$ uncertainty in the $^{8}$B solar neutrino flux prediction. Additional background contributions are included with their respective uncertainties: $4.8\%$ for accidental coincidences (AC), $50\%$ for neutron-induced background, and $100\%$ for the subleading electron-recoil (ER) background. These priors are incorporated in the statistical analysis via Gaussian penalty terms in the $\chi^2$ function. Finally, for each new physics parameter in $\overrightarrow{\mathcal{S}}$ ($g_X$ and $m_X$), the $\chi^2$ function is marginalized over all nuisance parameters.

As discussed in the Introduction (Section~\ref{Sec:Introduction}), we consider two distinct scenarios:
\begin{itemize}
    \item 
In scenario (i), neutrinos interact with nuclei via the SM \cevns~cross section (see Eq.~\eqref{Eq.CEvNS_SM_xSec}), while DM couples to nuclei through a light scalar, vector or axial-vector mediator (Eqs.~\eqref{eq:DM_nuclei_xsec_newphysicsS}–\eqref{eq:DM_nuclei_xsec_newphysicsA}).
\item 
In scenario (ii), on the contrary, we assume effective SI or SD DM–nucleon interactions, while allowing neutrinos to exhibit BSM interactions with nuclei mediated by a light mediator, cf. Eqs.~\eqref{Eq.SM_plus_S_CEvNS_xSec}–\eqref{Eq.SM_plus_A_CEvNS_xSec}.
\end{itemize}

\begin{figure}[!ht]
    \centering
        \includegraphics[width=0.49\linewidth]{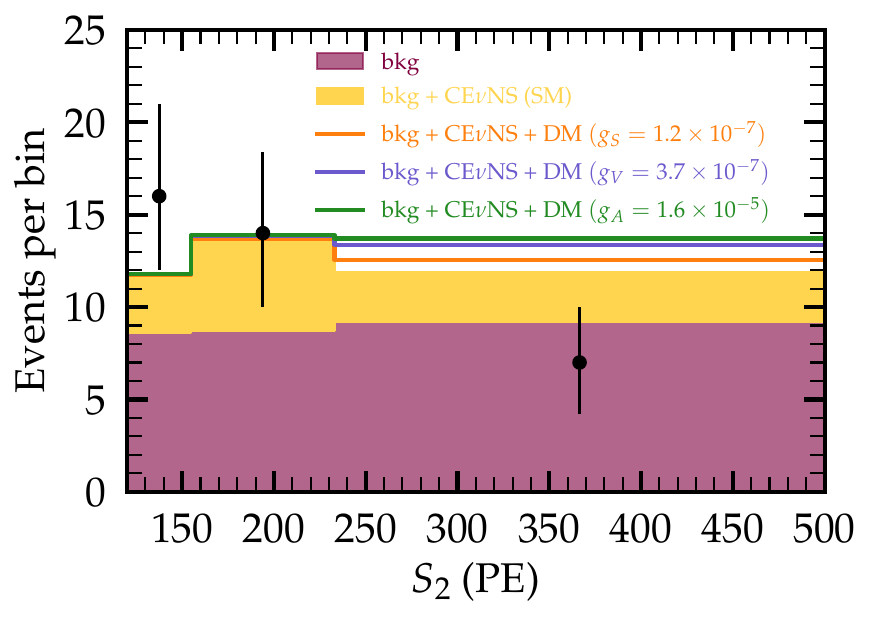}
        \includegraphics[width=0.49\linewidth]{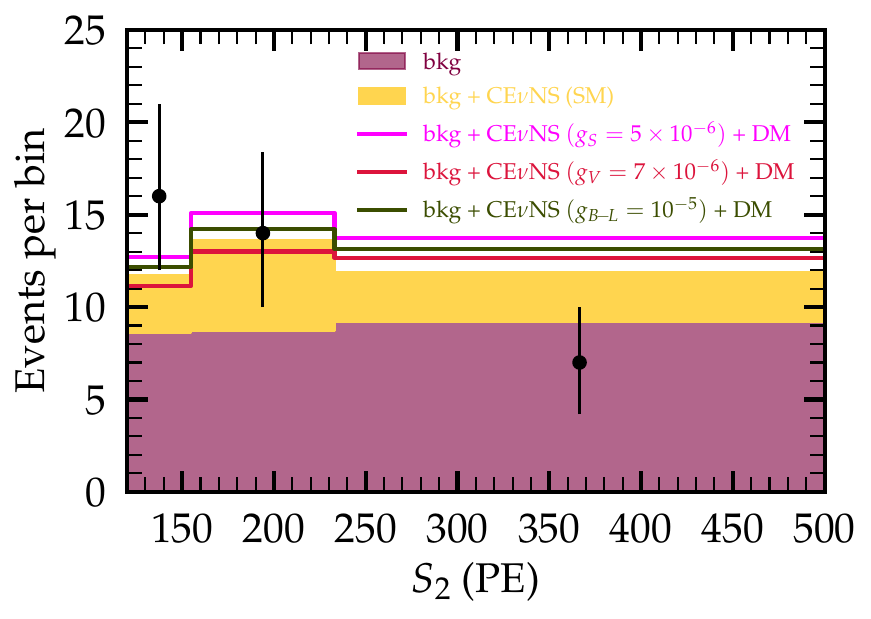}
    \caption{Expected nuclear recoil spectra for XENONnT compared with data from~\cite{XENON:2024ijk,XENON:2024hup}. Maroon histograms indicate the total background. {\bf Left panel}: scenario (i), showing the SM \cevns\ contribution from $^8$B neutrinos (yellow), and together with benchmark DM–nucleus interactions mediated by light scalar (orange), vector (blue), or axial-vector (green) particles. For the BSM DM benchmarks, we fix $m_\chi = 100~\mathrm{GeV}$ and $m_X = 1~\mathrm{MeV}$. {\bf Right panel}: scenario (ii), displaying the SM \cevns\ signal from $^8$B neutrinos (yellow), and benchmarks for \cevns\ in the presence of additional light scalar and vector mediators, together with a DM component corresponding to a point-like SI interaction with $m_\chi = 13~\mathrm{GeV}$ and $\left.\sigma_{\chi\mathfrak{n}}^{\mathrm{SI}}\right|_{\qtransfer^{2}\to 0} = 10^{-47}~\mathrm{cm^2}$. In all BSM \cevns\ cases, we fix $m_X = 1~\mathrm{MeV}$.}
    \label{fig:XENONnT_Spectra}
\end{figure}

We show in Fig.~\ref{fig:XENONnT_Spectra} the expected event rates per recoil-energy bin together with the XENONnT data points from~\cite{XENON:2024ijk,XENON:2024hup} and their associated error bars. The maroon  histograms represent the sum of AC, neutron-induced, and ER backgrounds.
The left panel corresponds to scenario (i) and shows the following cases:
(1) background plus the SM \cevns\ contribution from $^8$B solar neutrinos (yellow);
(2) background plus SM \cevns\ and a DM–nucleus interaction mediated by a light scalar (orange), with $g_S = 1.2\times 10^{-7}$, $m_S = 1~\mathrm{MeV}$, and $m_\chi = 100~\mathrm{GeV}$;
(3) background plus SM \cevns\ and a DM–nucleus interaction mediated by a light vector (blue), with $g_V = 3.7\times 10^{-7}$, $m_V = 1~\mathrm{MeV}$, and $m_\chi = 100~\mathrm{GeV}$;
(4) background plus SM \cevns\ and a DM–nucleus interaction mediated by a light axial-vector (green), with $g_A = 1.6\times 10^{-5}$, $m_A = 1~\mathrm{MeV}$, and $m_\chi = 100~\mathrm{GeV}$.
In this scenario, the mediator masses and couplings are selected to satisfy the bounds derived from our analysis of the XENONnT data, which will be presented in the next section.
The right panel corresponds instead to scenario (ii) and displays the following cases:
(1) background plus the \cevns\ contribution from $^8$B solar neutrinos assuming the SM cross section (yellow);
(2) background plus the \cevns\ contribution in the presence of a light scalar mediator (magenta), using $g_S = 5\times 10^{-6}$ and $m_S = 1~\mathrm{MeV}$;
(3) background plus the \cevns\ contribution including a light universal vector mediator (red), with $g_V = 7\times 10^{-6}$ and $m_V = 1~\mathrm{MeV}$;
(4) background plus the \cevns\ contribution including a light $B\!-\!L$ vector mediator (dark green), using $g_{B\!-\!L} = 1\times 10^{-5}$ and $m_V = 1~\mathrm{MeV}$.
In cases (2)–(4) in the right panel, we also include a DM component corresponding to a point-like SI interaction with $m_\chi = 13~\mathrm{GeV}$ and $\left.\sigma_{\chi\mathfrak{n}}^{\mathrm{SI}}\right|_{\qtransfer^{2}\to 0} = 10^{-47}~\mathrm{cm^2}$.
In the case of \cevns~with BSM interactions, we choose values of mediator masses and couplings satisfying bounds from a combined XENONnT and PandaX-4T nuclear-recoil analysis~\cite{DeRomeri:2024iaw} (though slightly above the more recent constraint from CONUS+~\cite{DeRomeri:2025csu} in the case of a universal vector mediator), for the sake of illustration.
 Note that even stronger limits on new interactions in the neutrino sector can in principle apply, especially from elastic neutrino-electron scattering data at Borexino~\cite{Coloma:2022umy}, CHARM-II~\cite{Bauer:2018onh}, and TEXONO~\cite{TEXONO:2009knm,Bauer:2018onh}, as well as from a combined analysis of PandaX-4T, XENONnT, and LZ electron-recoil data~\cite{A:2022acy,DeRomeri:2024dbv}. However, we are focusing here on nuclear recoils, hence one may assume different couplings of the new mediator to neutrinos, electrons and quarks (although not easy to motivate from a gauge-invariance point of view).

Finally, we refer the reader to Appendix~\ref{sec:Appendix_A} for more details on the \cevns~and DM spectral shapes in presence of light mediators.

%%%%%%%%%%%%%%%%%%%%%%%%%%%%%%%%%%%%%%%%%%%%%%%%%%%%%%%%%%%%%%%%%%%%%%%%%%%%%%%%
\section{\label{Sec:Neutrino_Fog} Estimation of the neutrino fog}
%%%%%%%%%%%%%%%%%%%%%%%%%%%%%%%%%%%%%%%%%%%%%%%%%%%%%%%%%%%%%%%%%%%%%%%%%%%%%%%%
The sensitivity frontier of DM direct detection experiments is characterized by the \textit{discovery limit}, i.e., the smallest DM–nucleon scattering cross section that can be probed before the signal becomes  indistinguishable from the neutrino background. From a statistical perspective, one approach to quantify this regime is through the concept of the so-called \textit{neutrino fog}~\cite{OHare:2021utq}. This definition is based on a derivative of the discovery limit as a function of the exposure.

To estimate the neutrino fog we first compute binned integrated spectra for DM-nuclei scattering and/or \cevns~events in the presence of a BSM mediator $X=S,V,A$, as

\begin{equation}
R_{\nu \mathcal{N} (\chi \mathcal{N})}^{X, i} = \int_{T_\mathcal{N}^i}^{T_\mathcal{N}^{i+1}} \d T_\mathcal{N} {\left[\frac{\d R}{\d T_\mathcal{N}}\right]}_{\nu \mathcal{N} (\chi \mathcal{N})}^X\,.
\end{equation}

Then, in order to quantify the discovery limits, we employ a profile likelihood ratio test~\cite{Cowan:2010js} which depends on the WIMP parameters ($m_\chi,\sigma_{\chi\mathfrak{n}}^\mathrm{SI~(SD)}$), the experiment exposure ($\varepsilon=t_\mathrm{run}m_\mathrm{det}$) and a set of nuisance parameters associated with the neutrino fluxes acting as background. Notice that among the relevant neutrino sources, only the $^8$B solar neutrinos produce a sufficiently large recoil rate to yield a detectable signal in current DM direct detection experiments, like XENONnT. Nonetheless, all other solar, atmospheric, DSNB, geoneutrino and reactor neutrino fluxes in principle enter the estimation of the neutrino fog, each through a different normalization factor $\mu_\mathrm{\nu}^{j}$ and a correspondent uncertainty $\sigma_\mathrm{\nu}^{j}$.
The likelihood is given as 

\begin{equation}
\label{Eq.Likelihood}
\mathcal{L}(m_\chi,\sigma_{\chi\mathfrak{n}}^{\mathrm{SI~
(SD)}},\varepsilon, \Phi^j_\mathrm{\nu})=\prod_i \mathcal{P}(R^i_{\mathrm{Exp}},R^i_{\mathrm{Obs}}) ~\prod_{j}~\mathcal{G}(\Phi_\mathrm{\nu}^{j},\mu_\mathrm{\nu}^{j},\sigma_\mathrm{\nu}^{j})\,,
\end{equation}
where $i$ denotes the energy bin while $j$ the neutrino flux.
Here, $\mathcal{P}(x,n)$ and $\mathcal{G}(x,\mu,\sigma)$ represent the Poisson and Gaussian probability distribution functions, respectively. This means that the observed number of events, $R_{\mathrm{Obs}}$, and the expected number of signal events, $R_{\mathrm{Exp}}$, are treated as Poisson-distributed random variables, while the nuisance parameters are modeled with Gaussian priors whose standard deviations correspond to the systematic uncertainties associated with the various neutrino fluxes.

To establish discovery limits, we define two hypotheses: the null hypothesis $H_0$, corresponding to a pure \cevns~background, and the alternative hypothesis $H_1$, in which a WIMP signal is present in addition to the \cevns~background. The likelihood function in Eq.~\eqref{Eq.Likelihood} is evaluated for both cases, yielding $\mathcal{L}_0$ for $H_0$ and $\mathcal{L}_1$ for $H_1$. In both scenarios, the total expected number of observed events is given by $R^i_{\mathrm{Obs}}=R^i_{\nu\mathcal{N}}(\Phi_\mathrm{\nu}^{j} = \mu_\mathrm{\nu}^{j})+R^i_{\chi\mathcal{N}}(m_\chi,\sigma_{\chi\mathfrak{n}}^{\mathrm{SI~(SD)}})$,
where $R^i_{\nu\mathcal{N}}$ and $R^i_{\chi\mathcal{N}}$ denote the neutrino–nucleus and WIMP–nucleus recoil rates, respectively. These depend on the WIMP mass $m_\chi$, the WIMP–nucleon cross section $\sigma_{\chi\mathfrak{n}}^{\mathrm{SI~(SD)}}$, the experimental exposure $\varepsilon$, and the neutrino flux normalization parameters $\Phi_\mathrm{\nu}^{j}$ fixed to their mean values.
Under the null hypothesis $\mathcal{L}_0$, the expected event count reduces to
$R^i_{\mathrm{Exp}}=R^i_{\nu\mathcal{N}}(\Phi_\mathrm{\nu}^{j})$,
where the neutrino flux normalizations are treated as a nuisance parameters. For the evaluation of $\mathcal{L}_1$, we employ the Asimov dataset approximation~\cite{Cowan:2010js}, setting $R^i_{\mathrm{Exp}} = R^i_{\mathrm{Obs}}$. In the regime of large statistics, this approximation provides accurate estimates of the discovery significance while ensuring computational efficiency.

Next, the discovery reach is quantified using the log-likelihood ratio test statistic, obtained by minimizing the negative log-likelihood functions over the nuisance parameters

\begin{equation}
q_0(m_\chi,\sigma_{\chi\mathfrak{n}}^{\mathrm{SI~(SD)}},\varepsilon) = -2\ln{\left[\frac{\mathcal{L}_0}{\mathcal{L}_1}\right]}_{\Phi_\nu^{j}~=~\hat{\Phi}_\nu^{j}} \,,
\end{equation}
where $\hat{\Phi}_\nu^{j}$ are the profiled maximum-likelihood estimators for each neutrino flux $\nu_j$. The neutrino fog is then characterized through the median discovery limit at a $3\sigma$ significance level. Specifically, for each DM mass and exposure, we identify the minimum DM-nucleon scattering cross section $\sigma_{\chi\mathfrak{n}}^{\mathrm{SI~(SD)}}$ that would allow for a $3\sigma$ discovery, i.e., satisfies $q_0 \geq 9$. The scaling behavior of  $ \sigma_{\chi\mathfrak{n}}^{\mathrm{SI~(SD)}}$ with respect to the exposure is encoded in the index~\cite{OHare:2021utq}

\begin{equation}
\label{Eq:n_index}
n = -\left(\frac{\d\ln\sigma_{\chi\mathfrak{n}}^{\mathrm{SI~(SD)}}}{\d\ln \varepsilon}\right)^{-1}. 
\end{equation}

In the Poisson-limited regime, where increasing exposure improves sensitivity as $\sigma_{\chi\mathfrak{n}}^{\mathrm{SI~(SD)}} \propto 1/\sqrt{\varepsilon}$, one finds $n = 2$. As the neutrino backgrounds become dominant and systematic uncertainties on the  neutrino fluxes start to limit the discovery potential, $n$ rises above 2, signaling the transition into the neutrino fog. Within this definition, the equivalent of the conventional neutrino floor can be identified with the contour $n = 2$ in the $(m_\chi, \sigma_{\chi\mathfrak{n}}^{\mathrm{SI~(SD)}})$ plane, marking  the regime with Poissonian background subtraction.  

We rely on the public code \texttt{NeutrinoFog}~\cite{OHare:2021utq,NuFog} to compute the neutrino fog, with tailored modifications in the computation of the event rates, so to account for the new interactions under consideration, and the updated values of the $\Phi_\nu^{^{8}\mathrm{B}}$ normalization and uncertainty, as inferred by XENONnT data.

\section{\label{Sec:Results} Results}

In this section, we present our results based on the latest XENONnT nuclear recoil data, focusing on the estimation of the neutrino fog and its potential modification in the presence of new physics. We examine two scenarios: (i) light mediators in the DM–nucleon interaction while neutrinos scatter with the target nuclei via the SM \cevns\ process, and (ii) effective SI or SD DM–nucleon interactions with new physics in the neutrino sector. In the former case, we derive 90\% confidence-level (CL) limits on scalar, vector, and axial-vector mediators and we evaluate their impact on the neutrino fog in both scenarios.

%----------------------------------------
\subsection{\label{Subsec:Results_CouplingMass} Constraints in the mediator coupling--mass plane}
%----------------------------------------
We start discussing limits on the new mediators from recent XENONnT nuclear recoil data~\cite{XENON:2024ijk,XENON:2024hup}. We focus on scenario (i), in which neutrinos scatter off nuclei via the SM \cevns~cross section (Eq.~\eqref{Eq.CEvNS_SM_xSec}), while DM interacts with nucleons via a light scalar,   vector, or axial-vector mediator (Eqs.~\eqref{eq:DM_nuclei_xsec_newphysicsS}–\eqref{eq:DM_nuclei_xsec_newphysicsA}).
While it is conventional to present DM direct detection limits in terms of the zero–momentum-transfer cross section $\left.\sigma^{\mathrm{SI~(SD)}}_{\chi\mathfrak{n}}\right|_{\qtransfer^2 \to 0}$, this parametrization obscures the interplay between mediator mass and momentum transfer, particularly in the light-mediator regime (see also~\cite{Bell:2023sdq,Fornengo:2011sz,DelNobile:2015uua}). To provide a more transparent picture, we first present our limits in the mediator coupling–mass parameter space, where the dependence on both $m_X$ and the coupling strength is explicit.
In this framework, for each new interaction $X$ ($=$ S, V or A), we define an effective coupling

\begin{equation}
g_X = \sqrt{g_\chi^X g_q^X}\, ,
\end{equation}
with $g_\chi^X$ and $g_q^X$ denoting the mediator couplings to DM and quarks, respectively, and assuming flavor-universal quark couplings $g_q^X = g_u^X = g_d^X$. This combination controls the overall normalization of the DM–nucleon scattering rate and provides a unified parameter for comparing different mediator scenarios.

\begin{figure}[t]
 \centering
 \includegraphics[width=\textwidth]{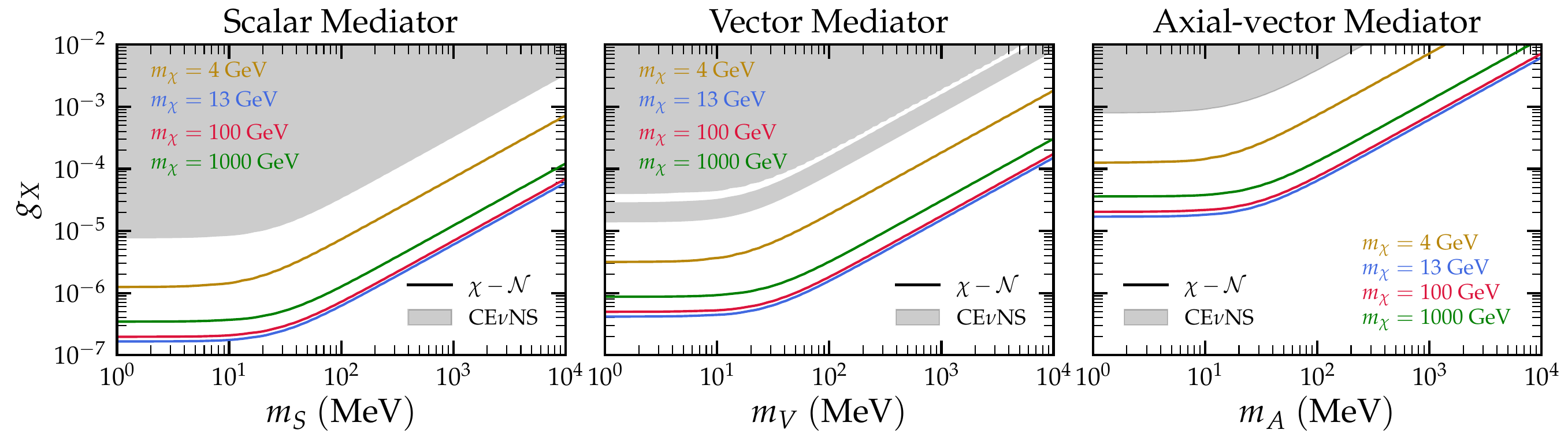}
 \caption{Constraints at 90\% CL in the new mediator coupling–mass plane, for scalar (left),  vector (center), and axial-vector (right) DM-nucleon interactions. The solid, colored curves show the XENONnT limits obtained when the new mediator couples to DM, for $m_\chi = 4$ (yellow), 13 (blue), 100 (red), and 1000~GeV (green). The shaded gray regions show the corresponding bounds when new physics appears only in the \cevns~sector (scenario~(ii), from~\cite{DeRomeri:2024iaw}; in this case, $g_X = \sqrt{g_\nu^X g_q^X}$). See text for more details.}
 \label{fig:BSM_Param_Space}
\end{figure}

Figure~\ref{fig:BSM_Param_Space} shows the resulting 90\% CL exclusion limits in the $(m_X,g_X)$ plane for scalar, vector, and axial-vector mediators, obtained from the analysis of XENONnT nuclear recoil data discussed in Sec.~\ref{subsec: XENONnT_analysis}. Constraints are shown for four representative DM masses, $m_\chi = 4,~ 13$, $100$, and $1000~\mathrm{GeV}$, illustrating how the sensitivity of XENONnT varies with the mediator and DM masses. The solid colored curves correspond to the limits derived in this work assuming that the new mediator couples only to DM (scenario (i)), while the gray shaded regions indicate the corresponding bounds when new physics 
 is present exclusively in the \cevns~sector (scenario~(ii)). For the latter case, we show the limits previously obtained in~\cite{DeRomeri:2024iaw} (see also complementary analyses in~\cite{Blanco-Mas:2024ale,AtzoriCorona:2025gyz}).
For completeness, let us note that other constraints from DM self-interactions (see, e.g.,~\cite{Bringmann:2016din,Tulin:2013teo}), neutrino self-interactions (see, e.g.,~\cite{Berryman:2022hds, Li:2023puz, Esseili:2023ldf, Ghosh:2024cxi}), or bounds on a new light force carrier coupling to nucleons, may also apply and strongly constrain some regions of the light-mediator parameter space. However, usually they constrain only one coupling direction in parameter space (e.g., the mediator-neutrino coupling or the mediator-DM coupling). In contrast, DM direct detection nuclear-recoil (and \cevns) data probe the product of mediator couplings to DM (or neutrinos) and quarks. The two types of constraints therefore probe different directions in parameter space and are largely complementary, making a direct comparison non-trivial.

The bounds on the mediator coupling and mass obtained when the new interaction is introduced in the DM sector are more than one order of magnitude stronger than the corresponding limits derived when the new interaction is placed in the neutrino sector. This difference arises because the region of parameter space in which light mediators in the neutrino channel produce an observable deviation from the SM \cevns\ spectrum requires substantially larger couplings than those relevant for WIMP–nucleon scattering mediated by the same particle. The difference becomes more pronounced for DM masses in the 10–100 GeV range, where the DM event rate is highest. For lighter or heavier DM, the expected rate decreases
due to the low nuclear-recoil efficiency at low masses and the declining DM number density at high masses. This explains why the 4 GeV and 1000 GeV contours are less constraining than those for 13 GeV and 100 GeV. This trend will reappear in the discussion of the conventional DM sensitivity plane, $\left(m_\chi, \sigma_{\chi\mathfrak{n}}^\mathrm{SI~(SD)}\right)$, in the following section. 
Across most viable regions of parameter space, the XENONnT neutrino signal therefore remains consistent with the SM prediction, and the resulting limits predominantly constrain DM–nucleon interactions rather than new physics in the neutrino sector. This hierarchy persists even in the (less motivated) scenario where neutrinos and DM couple with equal strength to the new mediator, and it remains valid up to coupling ratios as large as $g_\nu^X \lesssim  40\, g_\chi^X$, assuming $m_\chi = 13$~GeV. For other DM masses, such as $m_\chi = 3.5$~GeV or $m_\chi = 1$ TeV, the maximum ratio $g_\nu^X / g_\chi^X$, for which the hierarchy still holds,   decreases and can approach unity, though it never falls below it.
Only for even larger values of $g_\nu^X$ do BSM effects in the neutrino sector become sizable, at which point the predicted \cevns\ rate starts to deviate appreciably from the SM expectation.

All in all, the coupling–mass representation of bounds highlights that the exclusion power depends not only on the overall interaction strength but also on the mediator mass, which determines the momentum-transfer dependence of the scattering amplitude. Constraints are significantly stronger when the mediator couples to DM, especially in the sub-100~MeV mediator mass range, where the typical momentum transfer becomes comparable to $m_X$. For larger mediator masses, the limits gradually converge toward the heavy-mediator (contact-interaction) regime. 
In this effective regime, we can obtain bounds on the SI $\left.\sigma^\mathrm{SI}_{\chi\mathfrak{n}}\right|_{\qtransfer^2 \to 0}$ or SD $\left.\sigma^\mathrm{SD}_{\chi\mathfrak{n}}\right|_{\qtransfer^2 \to 0}$ cross section versus the DM mass. We will show the inferred upper limits as orange regions in the panels of Figs.~\ref{fig:fog_SM} and~\ref{fig:fog_caseii}, in the following section. They are in good agreement with the official results from the XENONnT Collaboration~\cite{XENON:2024hup, XENON:2025vwd}.

%----------------------------------------
\subsection{\label{Subsec:Results_neutrino fog}
Implications for the neutrino fog}
%----------------------------------------

\begin{figure}[!ht]
  \centering
  \includegraphics[width=0.49\textwidth]{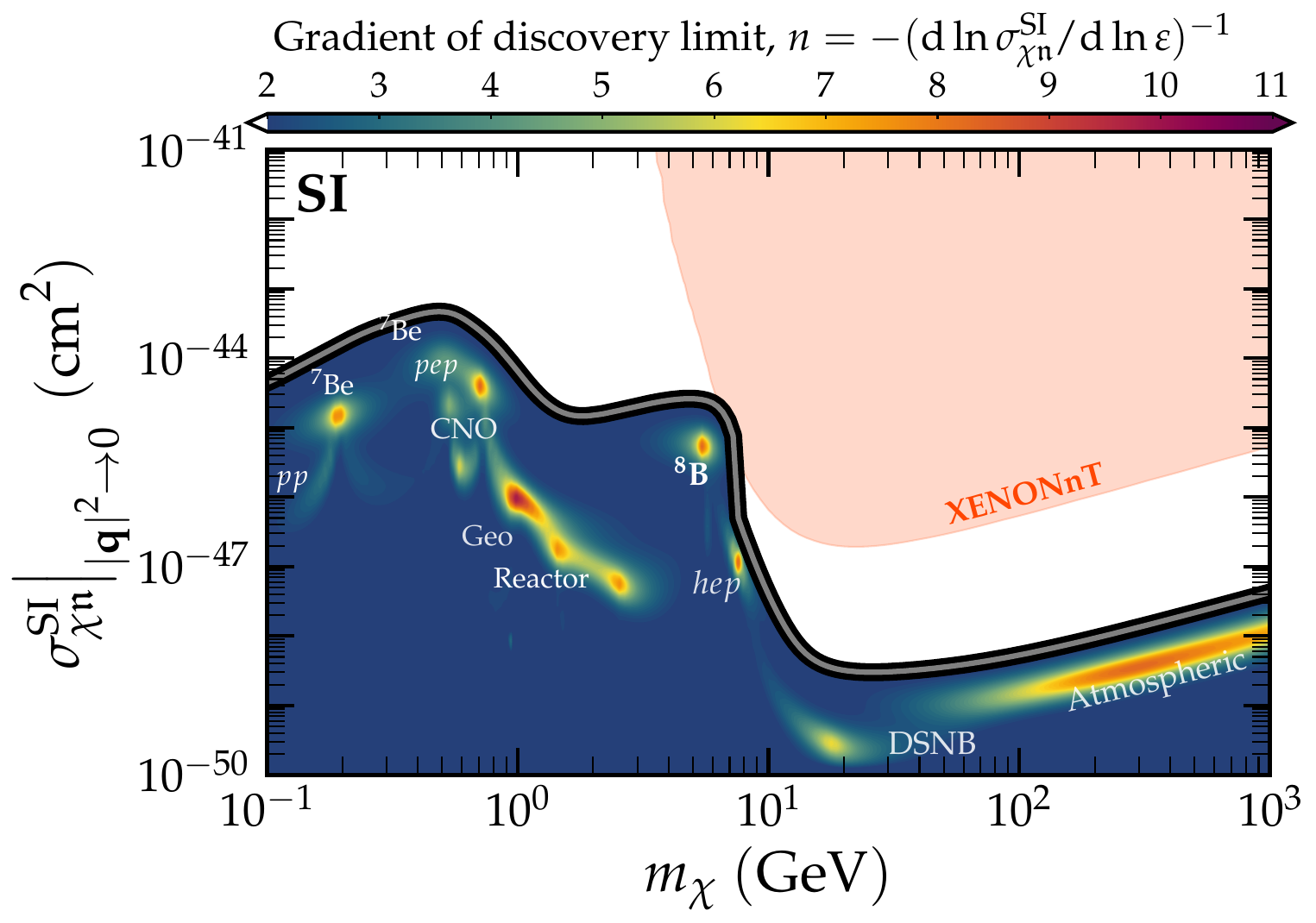}
    \includegraphics[width=0.49\textwidth]{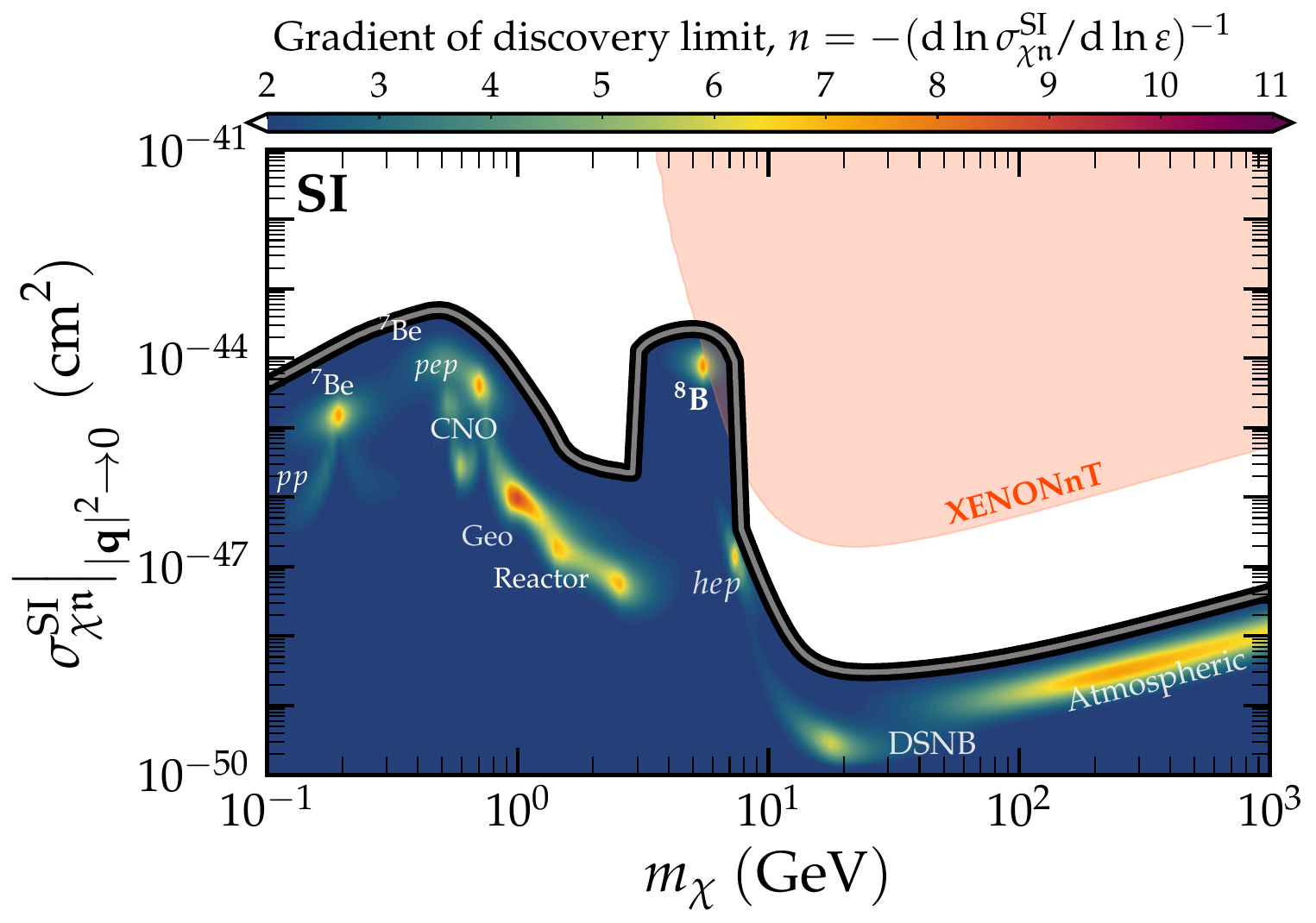}
    %%%%%%%%%%%%%%%%%
      \includegraphics[width=0.49\textwidth]{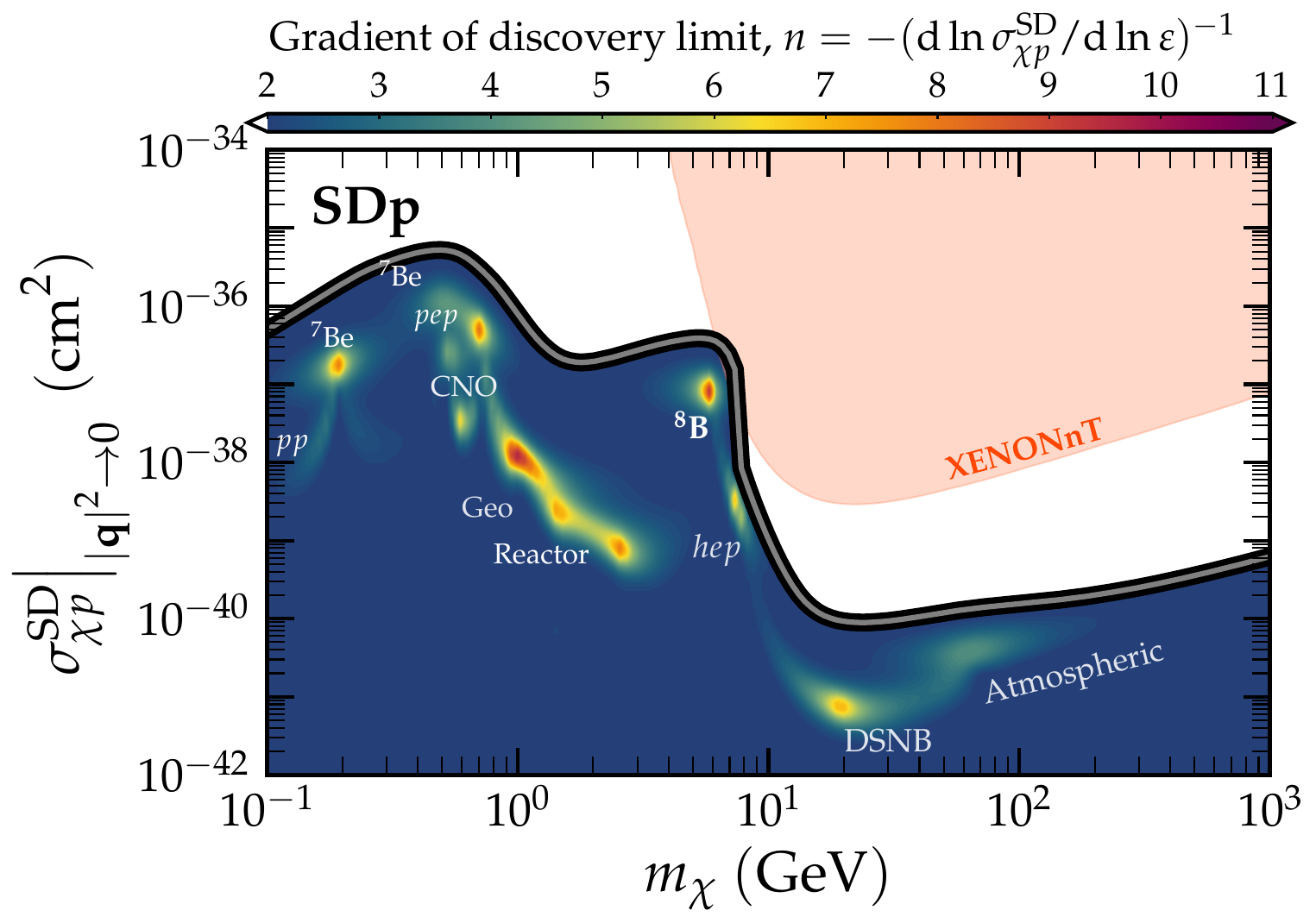}
    \includegraphics[width=0.49\textwidth]{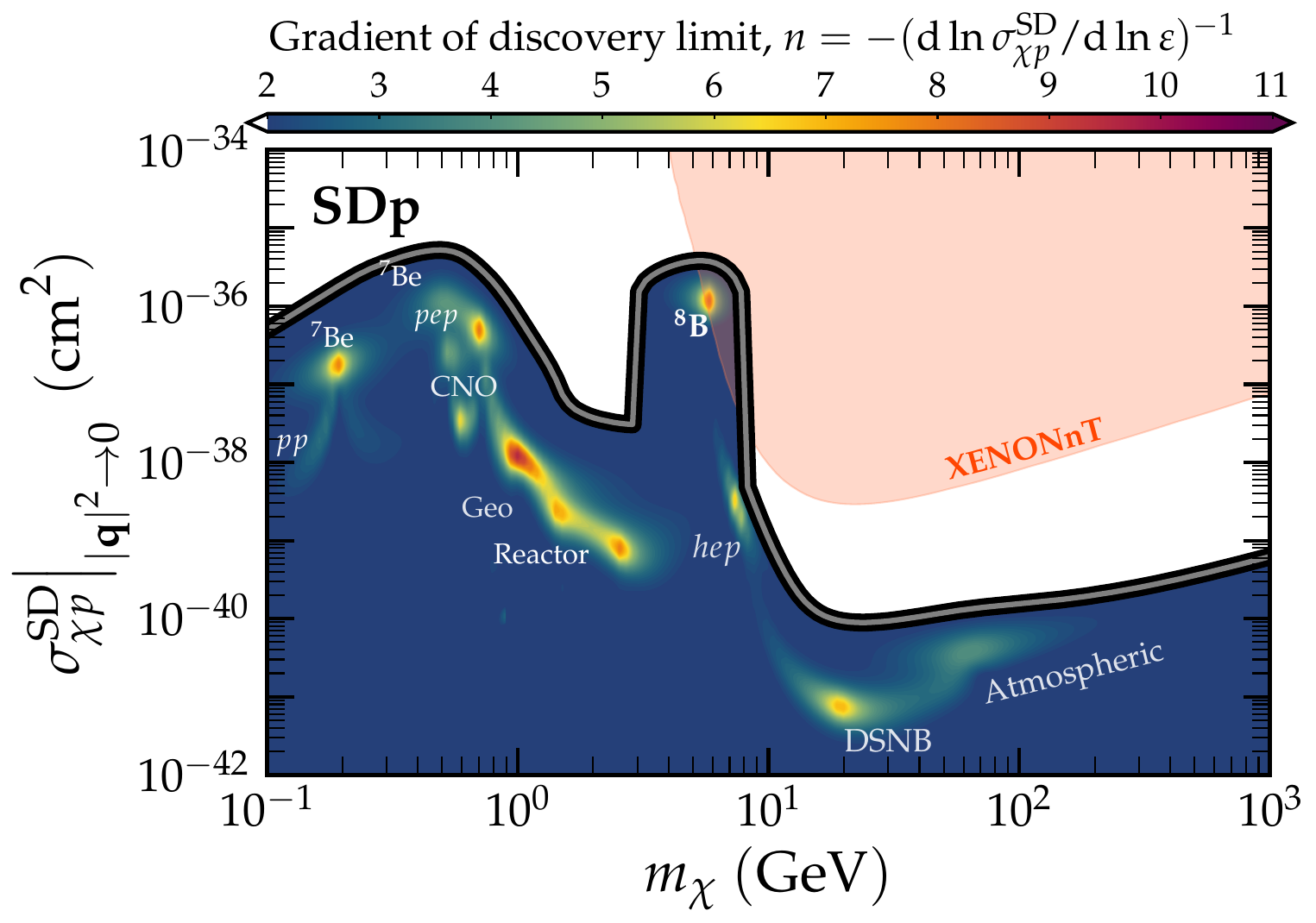}
    %%%%%%%%%%%%%%%%%%%%
      \includegraphics[width=0.49\textwidth]{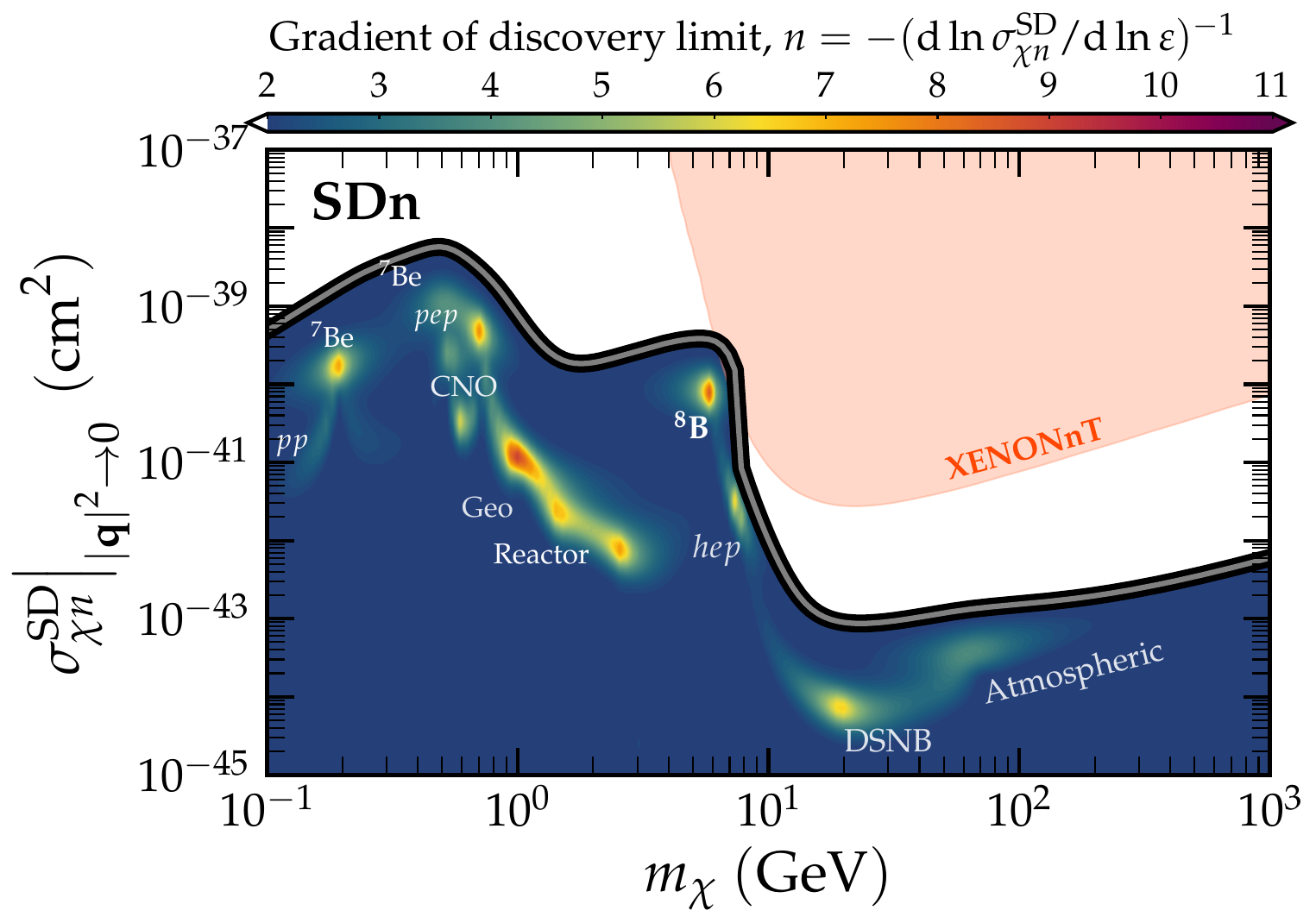}
    \includegraphics[width=0.49\textwidth]{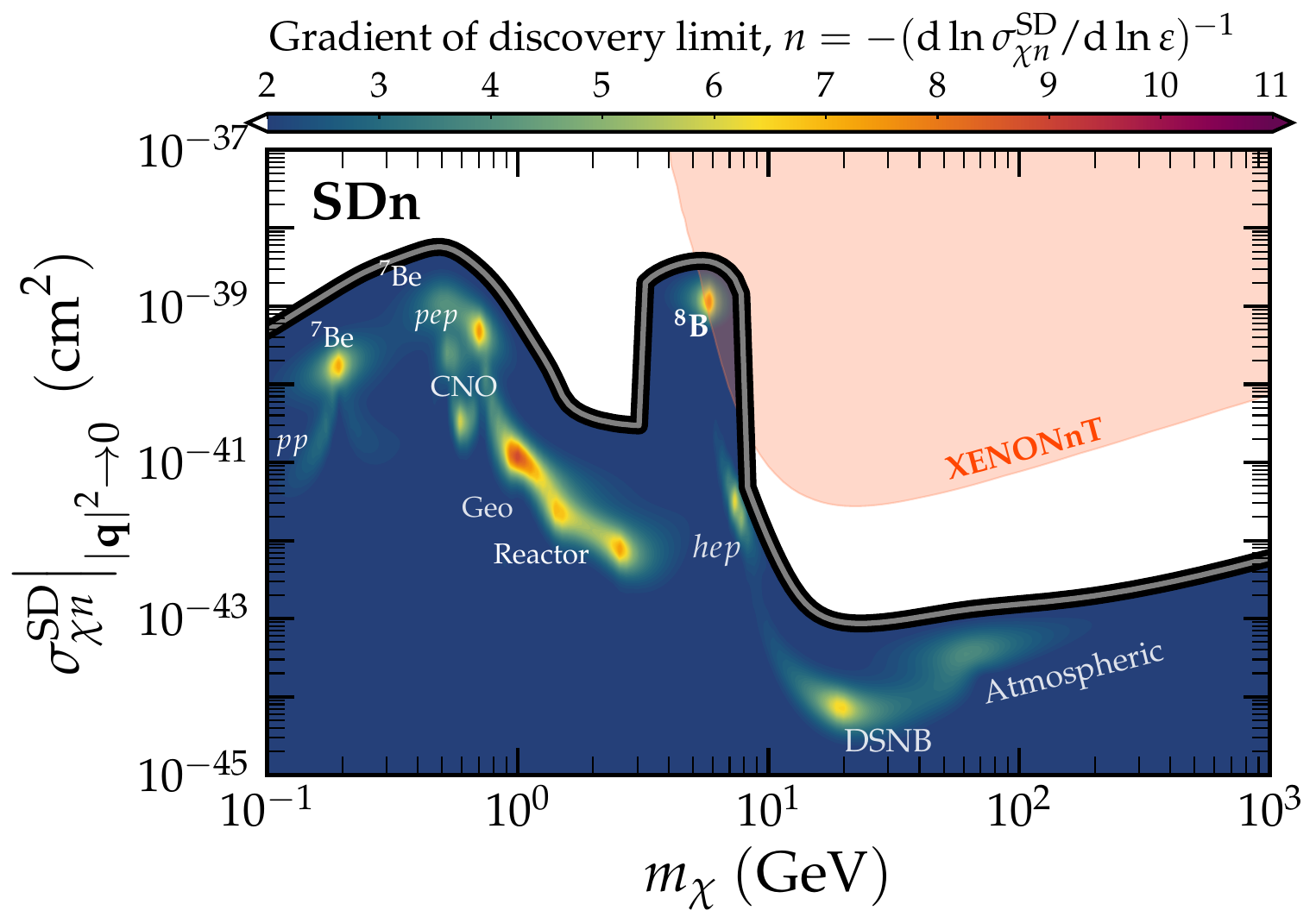}
  \caption{90\% CL upper limit (orange shaded areas) on the SI DM–nucleon (upper row), SD DM-proton (middle row) and SD DM-neutron (bottom row) cross section derived from our analysis of XENONnT data. {\bf Left panels}: standard definition of the neutrino fog with $^8$B flux normalization and uncertainty from SNO~\cite{Baxter:2021pqo,SNO:2011hxd}. {\bf Right panels}: neutrino fog with $^8$B flux normalization and uncertainty from the recent result of XENONnT~\cite{XENON:2024ijk}.}
  \label{fig:fog_SM}
\end{figure}

We now examine how the neutrino fog changes in light of recent XENONnT nuclear recoil data.
We begin our discussion by showing in Fig.~\ref{fig:fog_SM}  (panels in the left column) the reference neutrino fog obtained under the assumption of SM \cevns\ interactions for neutrinos and an effective DM–nucleon cross section, either SI (upper row) or SD with protons (middle row) or neutrons (bottom row). Recall that, given the definition in~\cite{OHare:2021utq}, also summarized in Sec.~\ref{Sec:Neutrino_Fog}, the color code of the neutrino fog corresponds to its \textit{opacity}, i.e., to the value of the index $n$ defined in Eq.~\eqref{Eq:n_index}. 
For this calculation, we adopt a recoil-energy window of $[10^{-4},200]~\mathrm{keV}_\mathrm{nr}$. Solar, atmospheric, and DSNB neutrino flux normalizations and uncertainties are implemented following the recommended values reported in~\cite{Baxter:2021pqo}; for $^8$B solar neutrinos specifically, we use $\mu_\nu^{^{8}\mathrm{B}} = 5.25\times10^6~\mathrm{cm^{-2}~s^{-1}}$ and $\sigma_\nu^{^{8}\mathrm{B}} = 4\%$~\cite{SNO:2011hxd}.
The right set of panels instead illustrates how the neutrino fog changes when using the $^8$B flux normalization and uncertainty extracted from the XENONnT measurement, namely $\mu_\nu^{^{8}\mathrm{B}} = 4.7\times10^6~\mathrm{cm^{-2}~s^{-1}}$ and $\sigma_\nu^{^{8}\mathrm{B}} = 63\%$~\cite{XENON:2024ijk}.
As expected, the significantly larger XENONnT uncertainty plays a relevant role in shifting the corresponding fog region towards higher cross sections, compared to the neutrino fog region obtained when the SNO measurement of the $^8$B flux is considered.  As explained before, a similar conclusion would have been obtained using the flux normalization and uncertainty obtained from PandaX-4T data~\cite{PandaX:2024muv}. With more data accumulated, we can expect this uncertainty to be reduced in the near future, and the fog contour to converge to the one predicted using other solar neutrino measurements.

\subsubsection{Scenario (i): light mediators in DM-nucleon interactions}
\label{Subsubsec:vfog_casei}

\begin{figure}[!htb]
  \centering
  \includegraphics[width=0.49\textwidth]{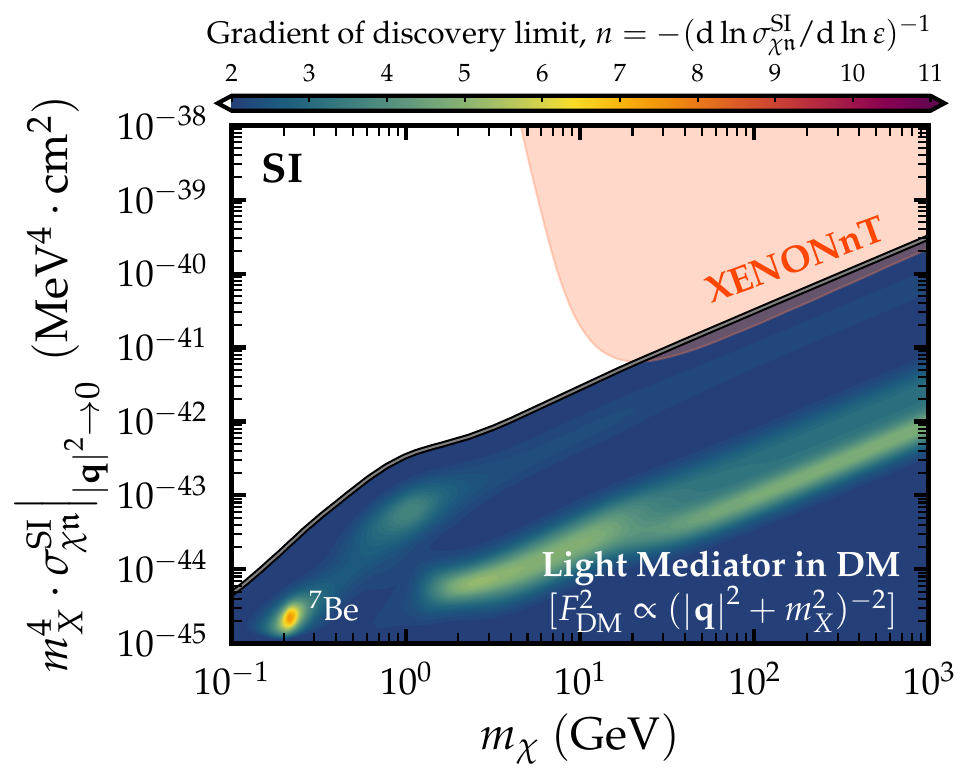}\\
    \includegraphics[width=0.49\textwidth]{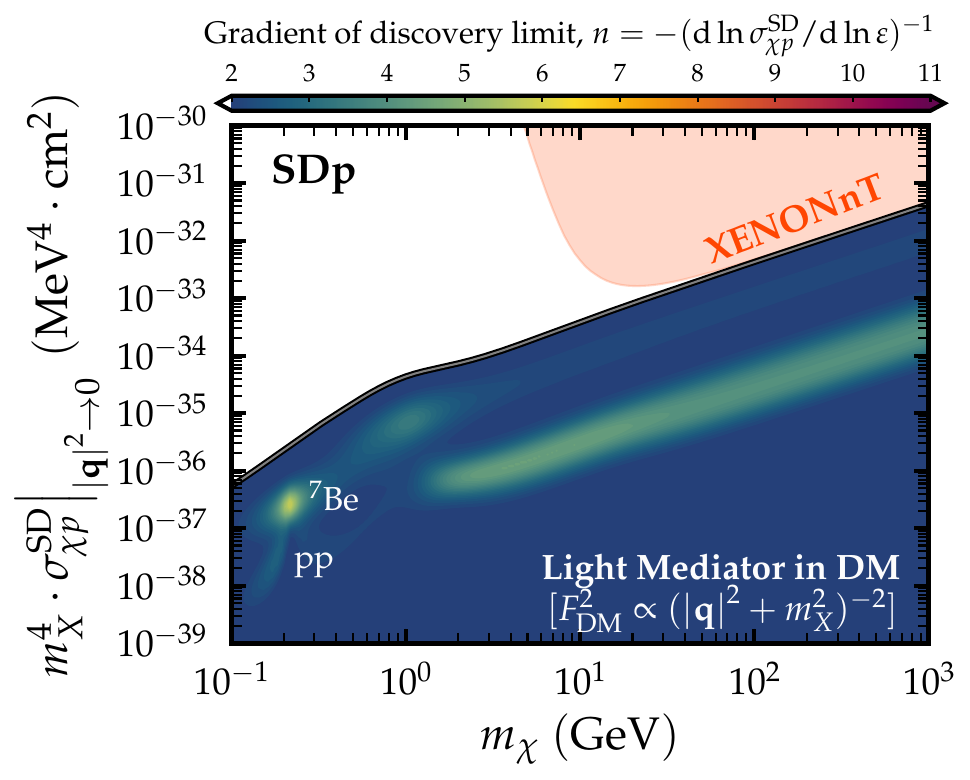}
        \includegraphics[width=0.49\textwidth]{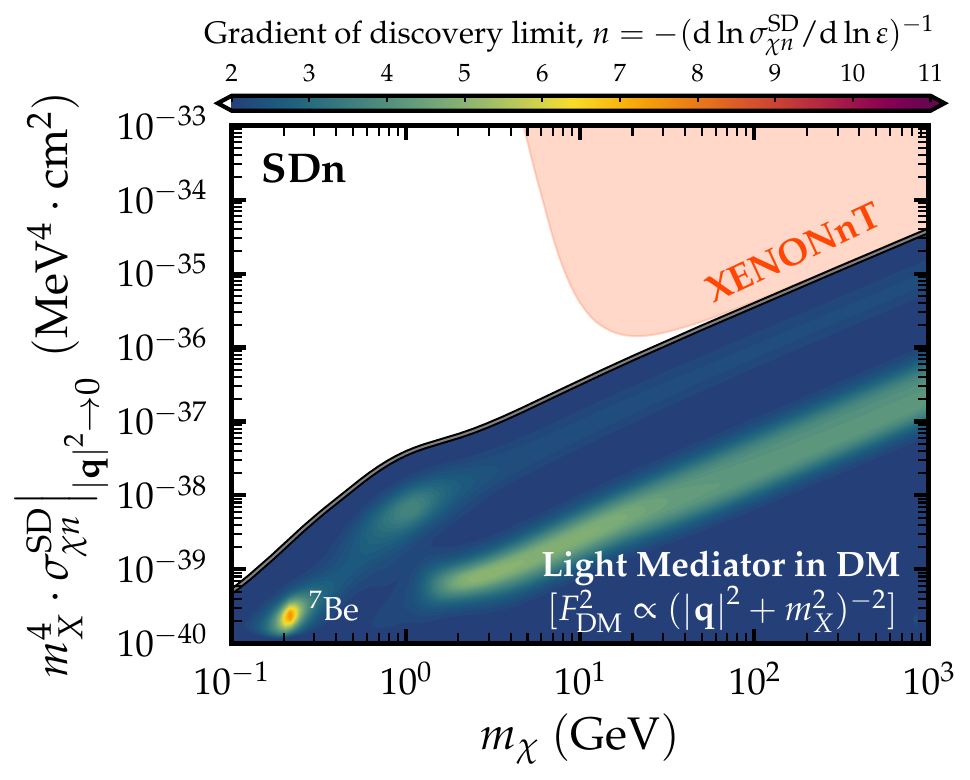}
  \caption{90\% CL upper limit (orange shaded area) on the SI (upper  panel), SD-proton (lower left panel), and SD-neutron (lower right panel) DM-nucleon cross section derived from our analysis of XENONnT data, under the assumption of DM interactions with a light mediator (scalar or vector for SI interactions, and axial-vector for the SD one). In all panels, the corresponding neutrino fog for scenario (i) is also shown.}
  \label{fig:fog_casei}
\end{figure}

We now continue by focusing on scenario (i) and examine in detail the impact of light scalar, vector, or axial-vector mediators in the DM sector on the neutrino fog. In this framework, neutrinos are assumed to interact via the SM \cevns~cross section. 
Figure~\ref{fig:fog_casei} displays the resulting constraints, from our analysis of XENONnT data, on the SI (left panel) and SD (right panel) DM–nucleon cross sections, shown as orange shaded regions. As a reminder, in this scenario, the differential DM scattering rate scales as $\propto F^{2}_\mathrm{DM}(\qtransfer^2) \sigma^\mathrm{SI~(SD)}_{\chi\mathfrak{n}} = \frac{m_X^4}{(\qtransfer^2+m_X^2)^2} \sigma^\mathrm{SI~(SD)}_{\chi\mathfrak{n}}$, leading to significantly modified bounds relative to the contact–interaction limit shown previously in Fig.~\ref{fig:fog_SM}. In particular, the dependence on both the mediator mass and the momentum transfer modify the inferred limits, especially in the light-mediator regime where $m_X \sim \qtransfer$.  Also in this case, we find good agreement with the official result, provided only for SI interactions, of the XENONnT Collaboration~\cite{XENON:2024hup}.

Under the same assumptions, we recompute the neutrino fog for a xenon target. The corresponding regions are also presented in Fig.~\ref{fig:fog_casei}. In this study, we keep the flux normalizations and uncertainties are implemented following the recommended values reported in~\cite{Baxter:2021pqo}, also for $^8$B solar neutrinos.
We observe a pronounced deformation of the fog, which is a direct consequence of the strong momentum-transfer dependence of the DM scattering rate. At low DM masses, the neutrino fog is significantly reduced: In this regime, the typical momentum transfers are small, and a light mediator enhances the DM–nucleon scattering rate proportionally to $\propto 1/\qtransfer^4$, yielding higher rates than the expected SM \cevns~contribution from solar neutrinos (notably from the $^8$B component) even at small $\sigma^\mathrm{SI~(SD)}_{\chi\mathfrak{n}}$. At large DM masses the opposite trend emerges: momentum transfers become larger, suppressing the DM rate for light mediators and reducing the relative impact of DM scattering compared to the neutrino background. This leads to a worsening of the discovery reach and a corresponding upward shift of the fog contour. Notice, also, that the opacity of the neutrino fog in this scenario is significantly reduced, implying an improved ability to distinguish between neutrino and DM event rates. The gradient of the discovery limit never exceeds $n=6$, not even in regions where neutrino fluxes are expected to be large. This means that smaller exposures will be enough to discriminate a DM signal from the neutrino background if DM actually interacts through light mediators, compared to the standard case with effective interactions. 
Unlike the SM-based neutrino fog shown in Fig.~\ref{fig:fog_SM}, the light-mediator scenario considered here does not give rise to well-defined regions of parameter space dominated by individual neutrino flux components, with the notable exception of the $^7$Be lines. Consequently, in Fig.~\ref{fig:fog_casei} we indicate only those regions where $^7$Be-induced \cevns~constitutes the sole background limiting the DM sensitivity. The broad yellow-shaded region extending over $m_\chi = 1$--$10^3$~GeV arises from the combined contributions of $pep$, $hep$, atmospheric, reactor, and geoneutrinos. Although the $^8$B solar neutrino flux does not produce a distinct shaded region, it significantly shifts the neutrino fog boundary upward toward the excluded parameter space.

\subsubsection{Scenario (ii): light mediators in \cevns}
\label{Subsubsec:vfog_caseii}

\begin{figure}[ht!]
  \centering
  \includegraphics[width=0.49\textwidth]{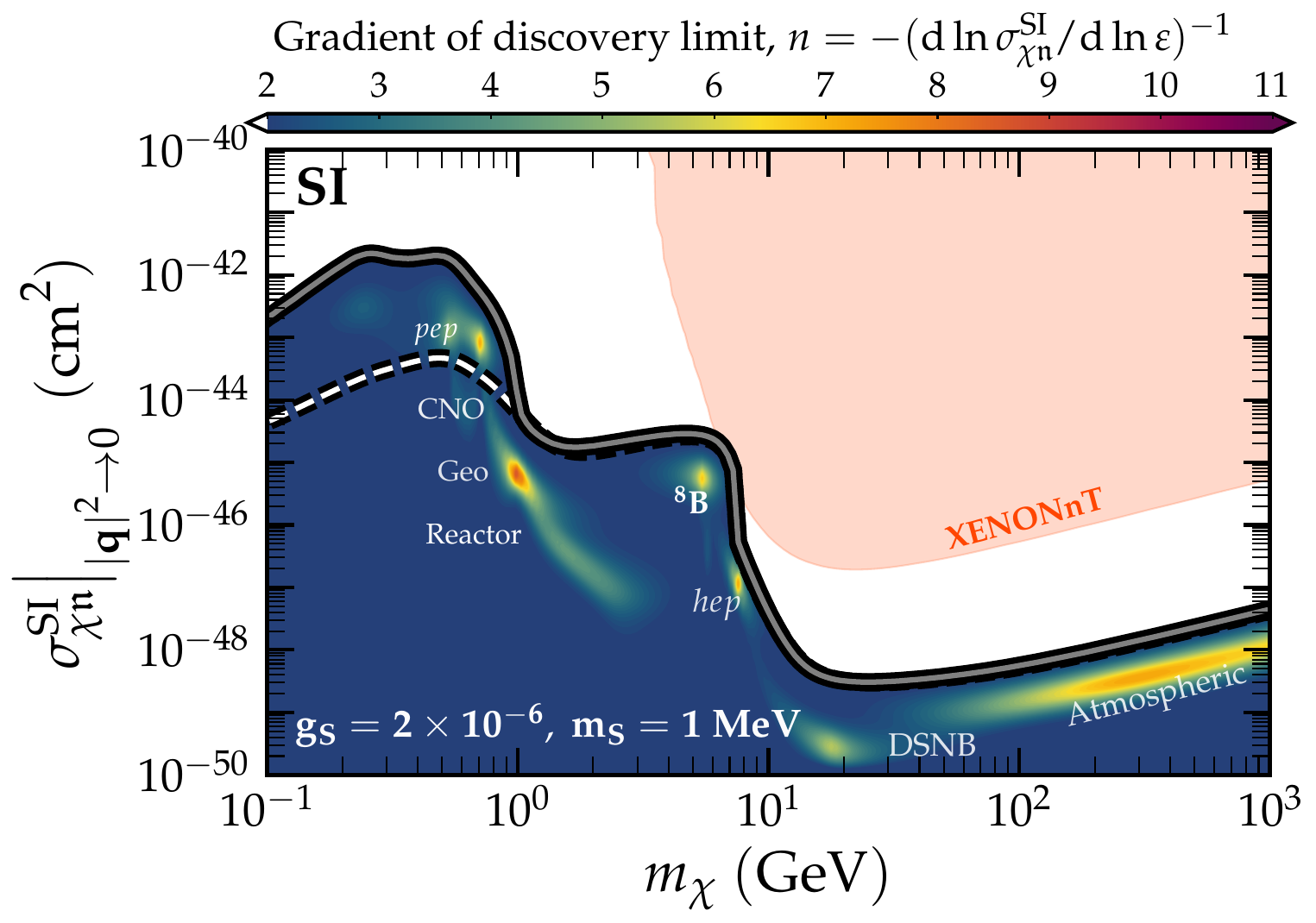}
  \includegraphics[width=0.49\textwidth]{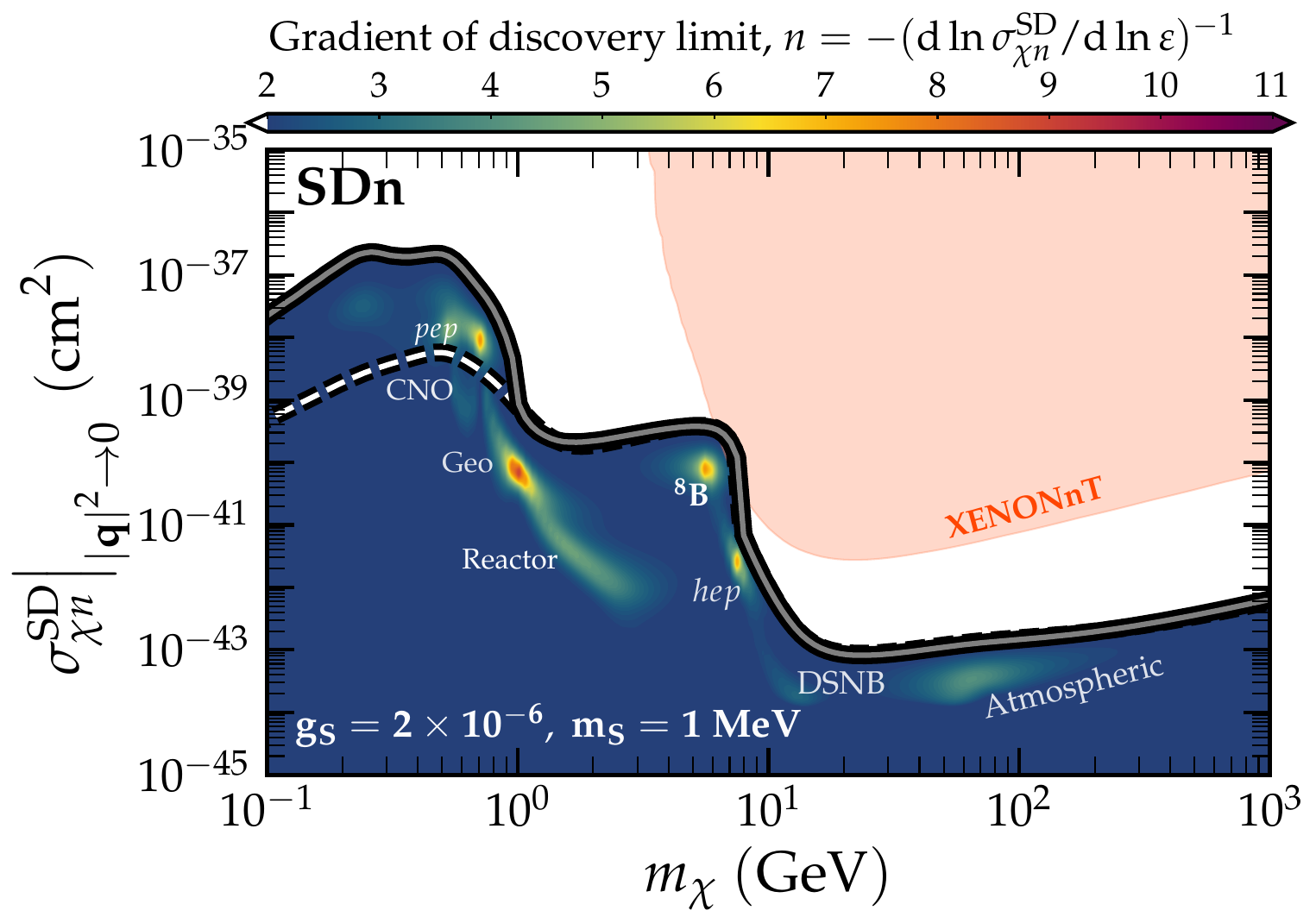}\\
  \includegraphics[width=0.49\textwidth]{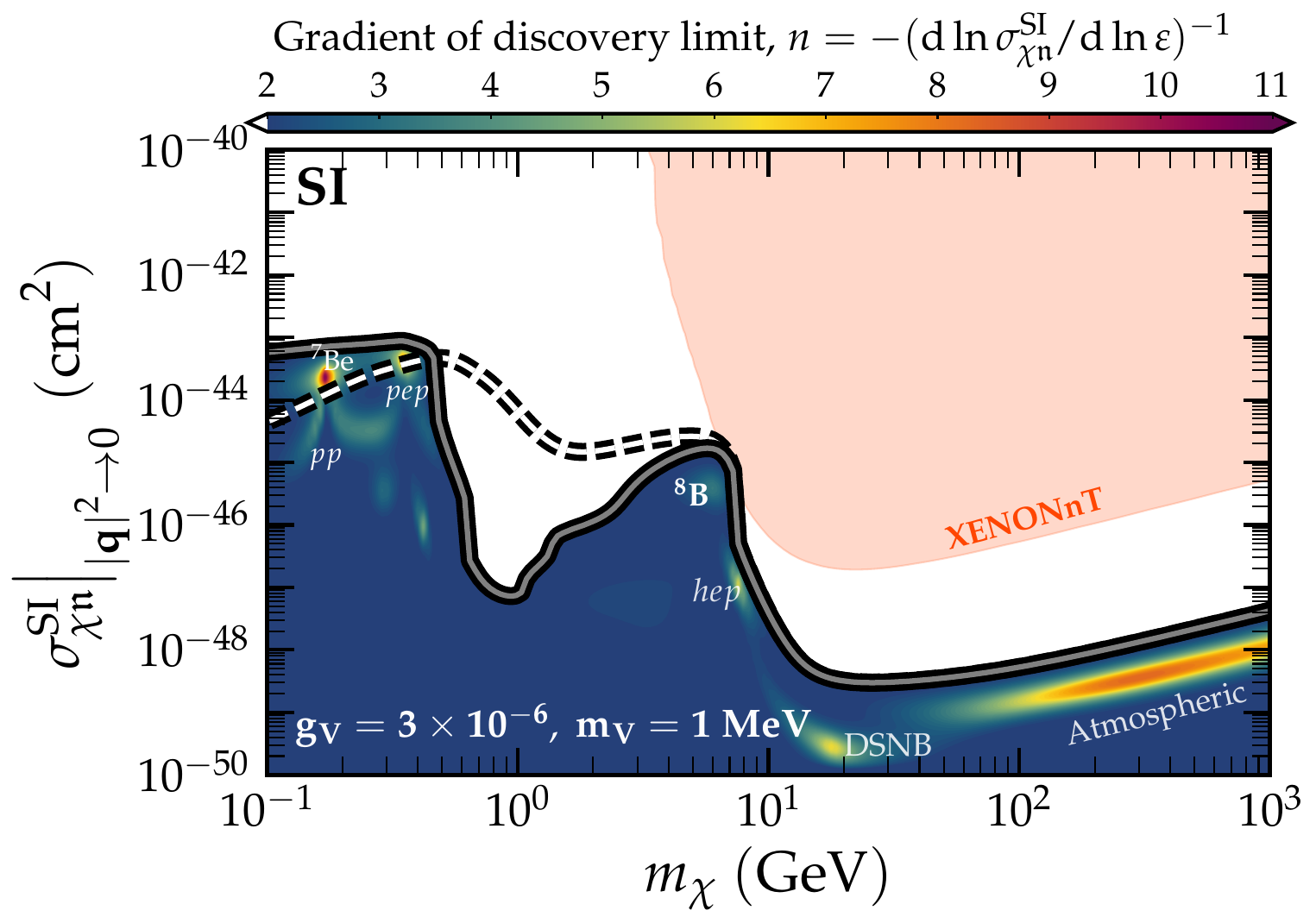}
  \includegraphics[width=0.49\textwidth]{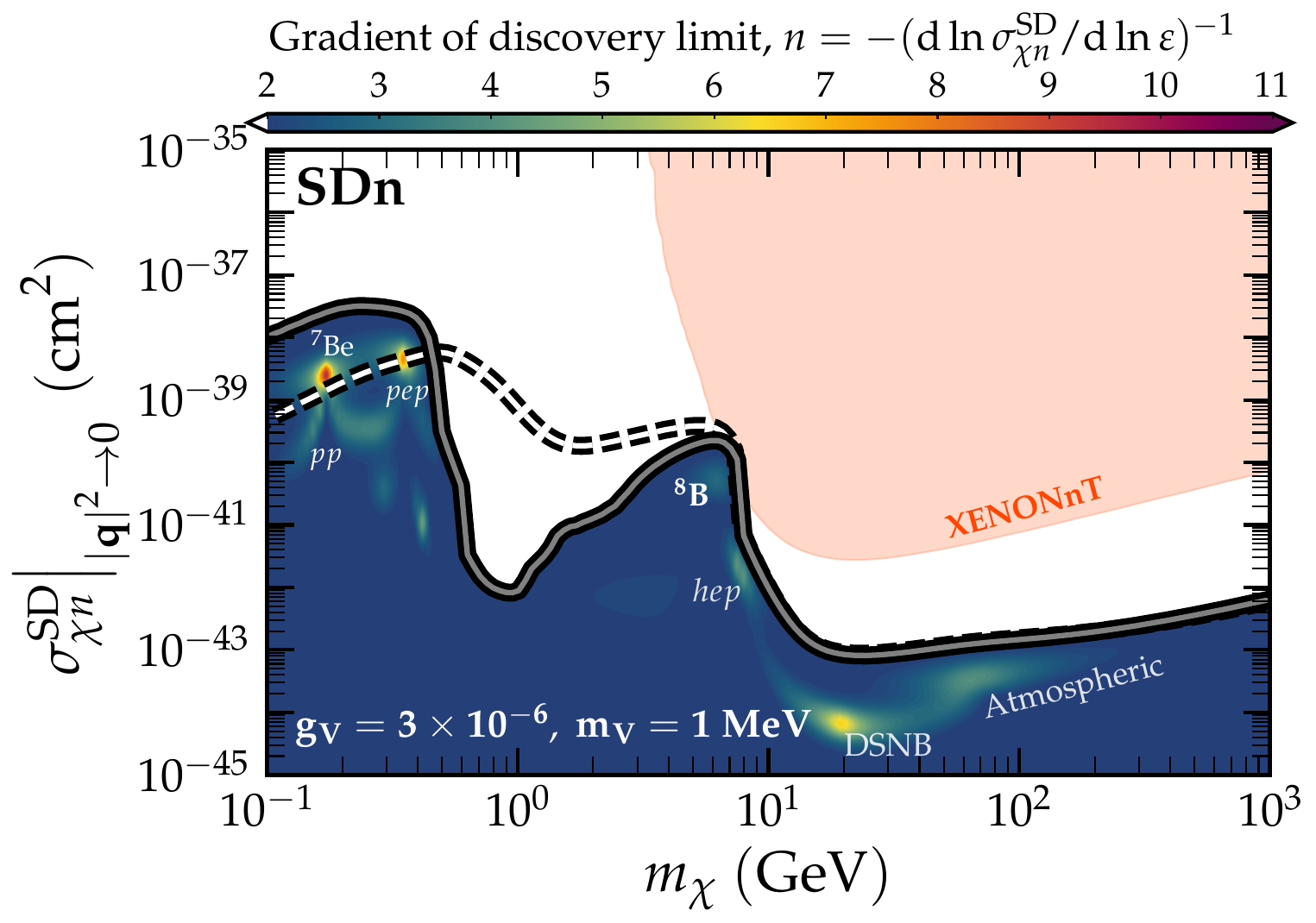}\\
  \includegraphics[width=0.49\textwidth]{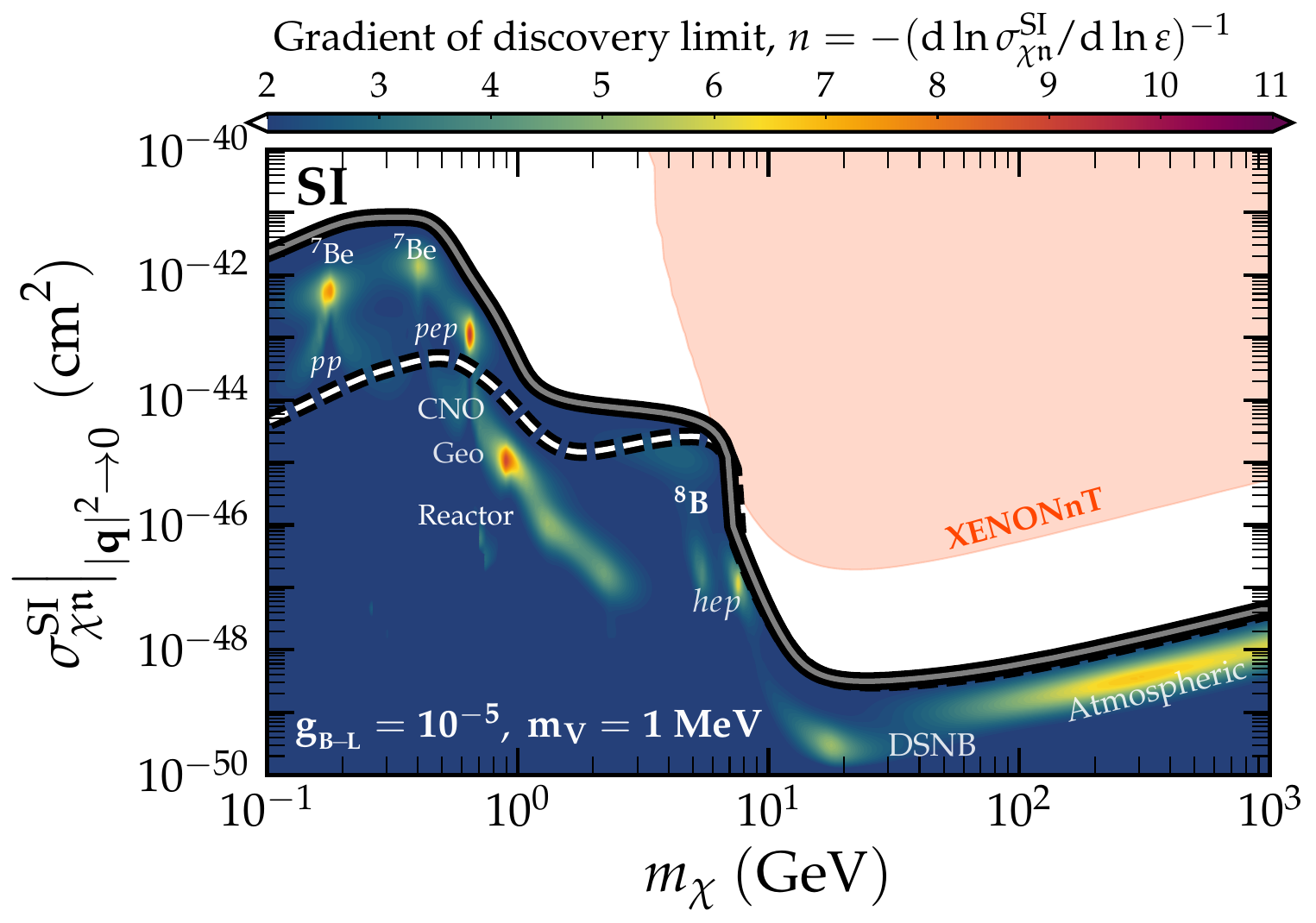}
  \includegraphics[width=0.49\textwidth]{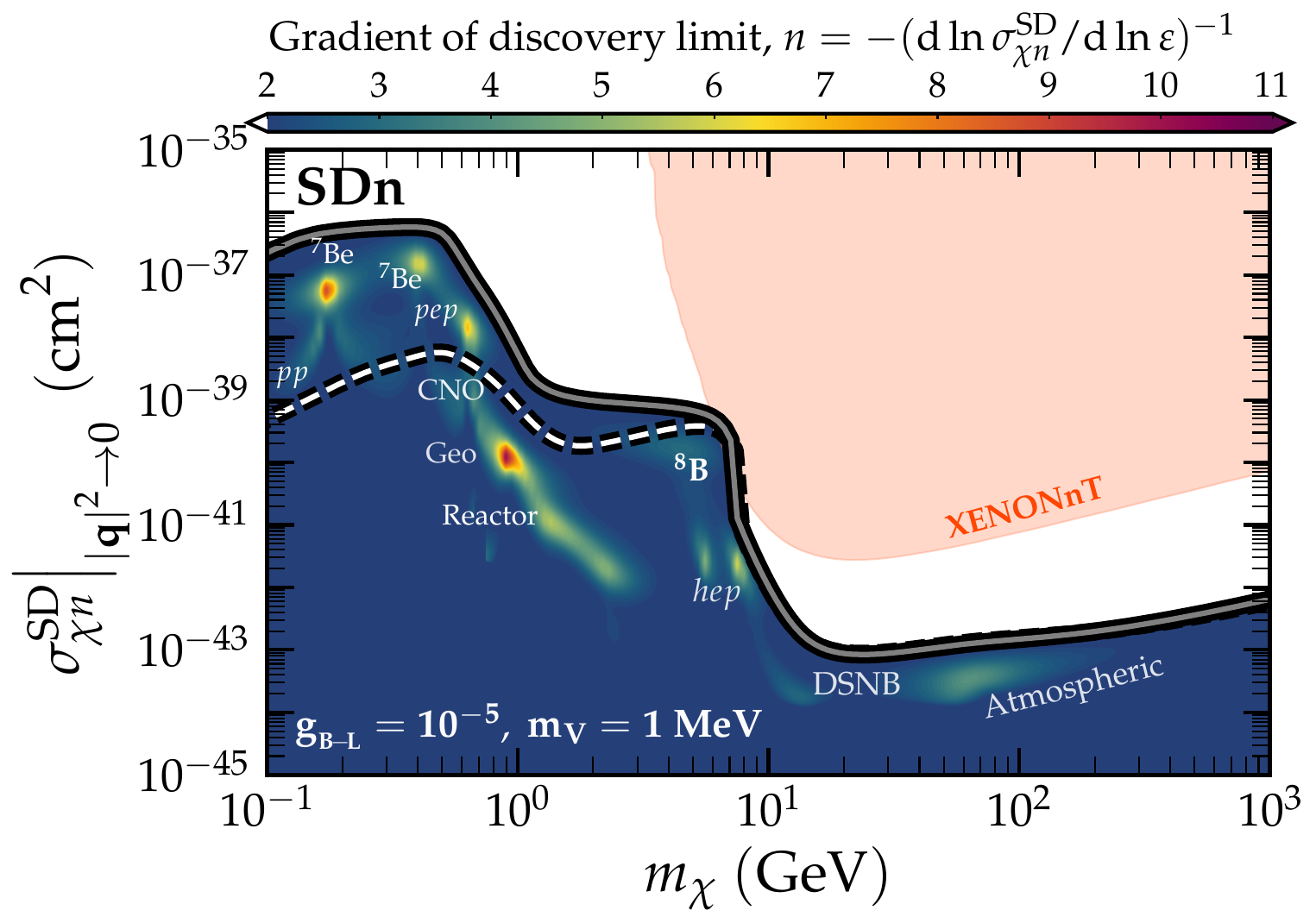}
  \caption{{\bf Left (right) panels}: neutrino fog for \cevns~interactions including a light scalar (top row), universal vector (middle row), or $B\!-\!L$ vector (bottom row) mediator, for SI DM-nucleon (SD DM-neutron) interactions. We fix the mediator masses to $m_{S} = m_{V} = m_{B\!-\!L} = 1~\mathrm{MeV}$ and the couplings to $g_S = 2\times10^{-6}$, $g_V = 3\times 10^{-6}$, and $g_{B\!-\!L} = 1\times10^{-5}$. The white dashed lines indicate the $n=2$ neutrino fog contours corresponding to the case in which neutrinos interact solely through the SM \cevns~process. }
  \label{fig:fog_caseii}
\end{figure}

We now turn to scenario (ii), where DM interacts with nucleons via an effective contact-type SI or SD interaction, while neutrinos scatter coherently off nuclei not only through the SM interaction but also through an additional scalar, vector, or axial-vector light mediator.
In this framework, the DM interaction corresponds to the heavy-mediator limit, in which the mediator has been integrated out, resulting in a constant cross section in the zero momentum-transfer regime. 
Consequently, the DM exclusion contours shown in Fig.~\ref{fig:fog_caseii} are identical to those obtained previously in the $(\left.\sigma_{\chi \mathfrak{n}}^{\mathrm{SI~(SD)}}\right|_{\qtransfer^2 \to 0}, m_\chi)$ plane (see Fig.~\ref{fig:fog_SM}).

On the other hand, the presence of new light mediators in the \cevns~process modifies the neutrino fog, potentially altering the discovery reach of upcoming DM direct detection experiments.
To quantify this effect, we define the following effective couplings

\begin{equation}
g_X \equiv \sqrt{g_\nu^X g_q^X}\, ,
\end{equation}
for each new interaction X ($=$ S, V or A),  where $g_\nu^X$ and $g_q^X$ denote the couplings of the mediator to neutrinos and quarks, respectively.

We illustrate in Fig.~\ref{fig:fog_caseii} the modified neutrino fog, in the presence of a new light scalar (upper row), universal vector (central row), or $B\!-\!L$ mediator (bottom row) in the \cevns~cross section, for both SI (left column) and SD (right columns) DM-nucleon interactions. 
We also show in all graphs the 90\% CL DM exclusion contours (orange shaded areas) previously obtained from the analysis of XENONnT nuclear recoil data. 
To compute the neutrino fog in this scenario, we use $\mu_\nu^{^{8}\mathrm{B}} = 5.25\times10^6~\mathrm{cm^{-2}~s^{-1}}$ and $\sigma_\nu^{^{8}\mathrm{B}} = 4\%$~\cite{SNO:2011hxd}, as well as all other flux normalizations and uncertainties as reported in~\cite{Baxter:2021pqo}. We also fix the new mediator mass and coupling as follows: $m_{S} =m_{V} = m_{B\!-\!L} = 1~\mathrm{MeV}$, $g_S = 2\times10^{-6}$, $g_V = 3\times10^{-6}$, and $g_{B\!-\!L} = 1\times10^{-5}$, values corresponding to solutions in the parameter space of the new mediator in agreement with the most recent bounds from \cevns~measurements, not only at DM direct detection experiments~\cite{DeRomeri:2024iaw}, but also at terrestrial ones such as COHERENT~\cite{DeRomeri:2022twg}, Dresden-II~\cite{Majumdar:2022nby,Coloma:2022avw}, and CONUS+~\cite{Chattaraj:2025fvx,DeRomeri:2025csu} (see discussion around Fig.~\ref{fig:XENONnT_Spectra}). In all panels, we further show as white dashed lines the $n=2$ fog contours corresponding to the case in which neutrinos interact solely through the SM \cevns~process, for the sake of comparison.

As illustrated in these figures, the inclusion of an additional light scalar or vector mediator in the \cevns~interaction modifies the morphology of the neutrino fog, most visibly through distortions of the $n=2$ contour. In the case of a light scalar mediator (top row), the new interaction adds incoherently to the SM amplitude, always increasing the total event rate. This enhancement is most pronounced at low DM masses, $m_\chi \lesssim 5$~GeV, where the relevant momentum transfers are small and the scalar contribution, scaling as $1/\qtransfer^4$, becomes dominant. The resulting increase in the predicted neutrino rate expands the neutrino fog region and worsens the DM discovery reach in this mass range.

For a light vector mediator with universal couplings (middle row), the new amplitude can interfere either constructively or destructively with the SM weak charge. At very low DM masses, the \cevns~ rate again gains a strong enhancement due to the behavior of the propagator, which scales as $1/(\qtransfer^2 + m_X^2)^2$ and is sizable in the solar--neutrino recoil range, leading to a degradation of the discovery potential similar to the scalar case. However, in the intermediate mass range, $1 \lesssim m_\chi \lesssim 10$~GeV, destructive interference becomes possible. This results in characteristic dips in the predicted recoil spectrum at specific recoil energies, where the SM and new--physics contributions partially cancel. Note that the exact position in parameter space of this cancellation is strongly dependent on the values of the mediator coupling and mass (see Appendix~\ref{sec:Appendix_A}). This local suppression of the neutrino rate translates into improved discovery reach in this DM mass interval, shifting down the fog contour. At higher DM masses, the fog is dominated by atmospheric neutrinos, which probe significantly larger momentum transfers ($\qtransfer \gg m_X$); in this regime, the BSM contribution is strongly suppressed, and the fog converges back to the standard expectation.

In the $B\!-\!L$ vector mediator scenario (bottom row), destructive interference cannot occur because the neutrino and quark charges are fixed by the gauge symmetry. Consequently, the \cevns~rate can only increase relative to the SM prediction. The neutrino fog therefore exhibits an expansion at low DM masses, driven once more by the enhancement of the rate at small momentum transfers, while remaining essentially unchanged at higher masses where the light mediator contribution essentially decouples.

Finally, we do not show results assuming an axial-vector interaction in \cevns~since, being suppressed due to its spin dependence, it does not lead to sizable modifications in the neutrino fog already shown in Fig.~\ref{fig:BSM_Param_Space}.

\section{\label{Sec:Conclusions} Conclusions}

The recent indications of \cevns\ from $^8$B solar neutrinos reported by XENONnT and PandaX-4T, while representing a major milestone for neutrino physics, have simultaneously increased the pressure on dark matter direct detection searches.  In parallel, observations of \cevns\ at terrestrial spallation sources and reactor experiments now allow a more accurate characterization of the irreducible neutrino background in direct detection experiments. An effective statistical framework to quantify the impact of this background, is the so-called \textit{neutrino fog}. 

In this work, we have investigated how the discovery reach of dark matter direct detection experiments is modified when either neutrinos or DM interact with nuclei via new light scalar, vector, or axial-vector mediators ($X=S,V,A$). Using the most recent results from XENONnT, as an example of current direct detection nuclear recoil data, we first derived updated bounds on mediator couplings and masses. These constraints prove to be significantly more stringent when the new mediator couples with DM than when it affects only neutrino interactions. We present our limits directly in the $(m_X, g_X)$ parameter space, rather than solely in the conventional $(m_\chi,~ \sigma_{\chi \mathfrak{n}}^\mathrm{SI~(SD)})$ plane, because this representation remains valid for both light and heavy mediators and naturally encodes the momentum-transfer dependence of the underlying interaction. We also translate our exclusion regions  into the standard $(m_\chi,~ \sigma_{\chi \mathfrak{n}}^\mathrm{SI~(SD)})$ plane and found them to be in good agreement with the official results from the XENONnT collaboration. We have also stressed that  the scalar and vector interactions result into identical exclusion curves when mapped into the $(m_\chi,~ \sigma_{\chi \mathfrak{n}}^\mathrm{SI})$ plane, further  motivating the presentation of constraints in terms of mediator masses and fundamental couplings as done in this work.

We then used these bounds to recompute the neutrino fog for each scenario and compared the results with the corresponding limits on spin-independent and spin-dependent DM–nucleon cross sections. We confirmed previous results in the literature and found that light mediators can substantially modify the profile of the neutrino fog. (i) When DM interacts through a light mediator, the neutrino fog is strongly reduced  at $m_\chi \lesssim 10$ GeV. The reason is that the DM–nucleus scattering rate is greatly enhanced at low momentum transfers, causing the DM signal to exceed the neutrino background more easily and thus improving the discovery potential; (ii) when neutrinos interact via an additional light scalar or a $B\!-\!L$ vector mediator, the discovery limits worsen at low DM masses. This is a direct consequence of the enhanced \cevns\ rates at small momentum transfers.
In contrast, for a light vector mediator with universal couplings, destructive interference with the SM \cevns\ amplitude is possible. This leads to an improvement of the discovery reach in the intermediate mass range, $1 \lesssim m_\chi \lesssim 10$ GeV, where dips in the neutrino spectrum suppress the background.

Overall, our results indicate that even in light of the latest XENONnT measurement, the presence of light mediators can significantly reshape the neutrino fog, altering the regions of parameter space where DM discovery becomes limited by neutrino backgrounds. Future detectors with directionality capabilities~\cite{Vahsen:2021gnb}, e.g., CYGNUS, are expected to provide new insights and possibly help overcoming the sensitivity limitations due to the neutrino background.\\

\noindent \textbf{Note Added:} Just before the submission of this work, the LZ Collaboration has also announced a 4.5$\sigma$ observation of \cevns~induced by $^8$B solar neutrinos and uploaded a preprint on their webpage~\cite{LZ-preprint}, and has appeared on the arXiv after our submission. If these new data were to be included in the present analysis, we expect conclusions to be qualitatively similar to those obtained using XENONnT data. The reason is that both Collaborations have measured the $^8$B neutrino flux with comparable $1\sigma$ uncertainties. A possible minor difference might arise from the fact that the LZ best-fit $^8$B flux normalization lies slightly below the theoretical prediction. In scenarios where new mediators interfere constructively with the SM amplitude, thus enhancing the \cevns~rate, this feature could lead to marginally stronger constraints.

%%%%%%%%%%%%%%%%%%%%%%%%%%%%%%%%%%%
\acknowledgments
%%%%%%%%%%%%%%%%%%%%%%%%%%%%%%%%%%%%
AM acknowledges financial support from the Government of India via the Prime Minister Research Fellowship (PMRF), ID: 0401970.
AM also gratefully acknowledges the hospitality of the AHEP group at IFIC, where a significant portion of the computations for this work was carried out.
VDR and DKP acknowledge financial support by the grant CIDEXG/2022/20 (from Generalitat Valenciana) and by the Spanish grants CNS2023-144124 (MCIN/AEI/10.13039/501100011033 and “Next Generation EU”/PRTR), PID2023-147306NB-I00, and CEX2023-001292-S (MCIU/AEI/ 10.13039/501100011033).

%%%%%%%%%%%%%%%%%%%%%%%%%%%%%%%%%%%%%%%%%%%%%%%%%%%
\appendix 
%%%%%%%%%%%%%%%%%%%%%%%%%%%%%%%%%%%%%%

%%%%%%%%%%%%%%%%%%%%%%%%%%%%%%%%%%%%%%%%%%%%%%%%%%%%
%----------------------------------------
\section{\label{sec:Appendix_A} Neutrino and dark matter scattering rates}
%----------------------------------------
In this Appendix, we present the expected \cevns\ and DM event rates for the two scenarios under considerations. In both cases, we assume a xenon target and fix an exposure of $1~\mathrm{ton\times year}$, while the binned spectra are evaluated assuming 200 logarithmically spaced bins in $[10^{-4},200]$ keV for \cevns\ rates and 227 bins in $[10^{-4},1400]$ keV for DM rates.

\begin{figure}[htb!]
  \centering
  \includegraphics[width=0.48\textwidth]{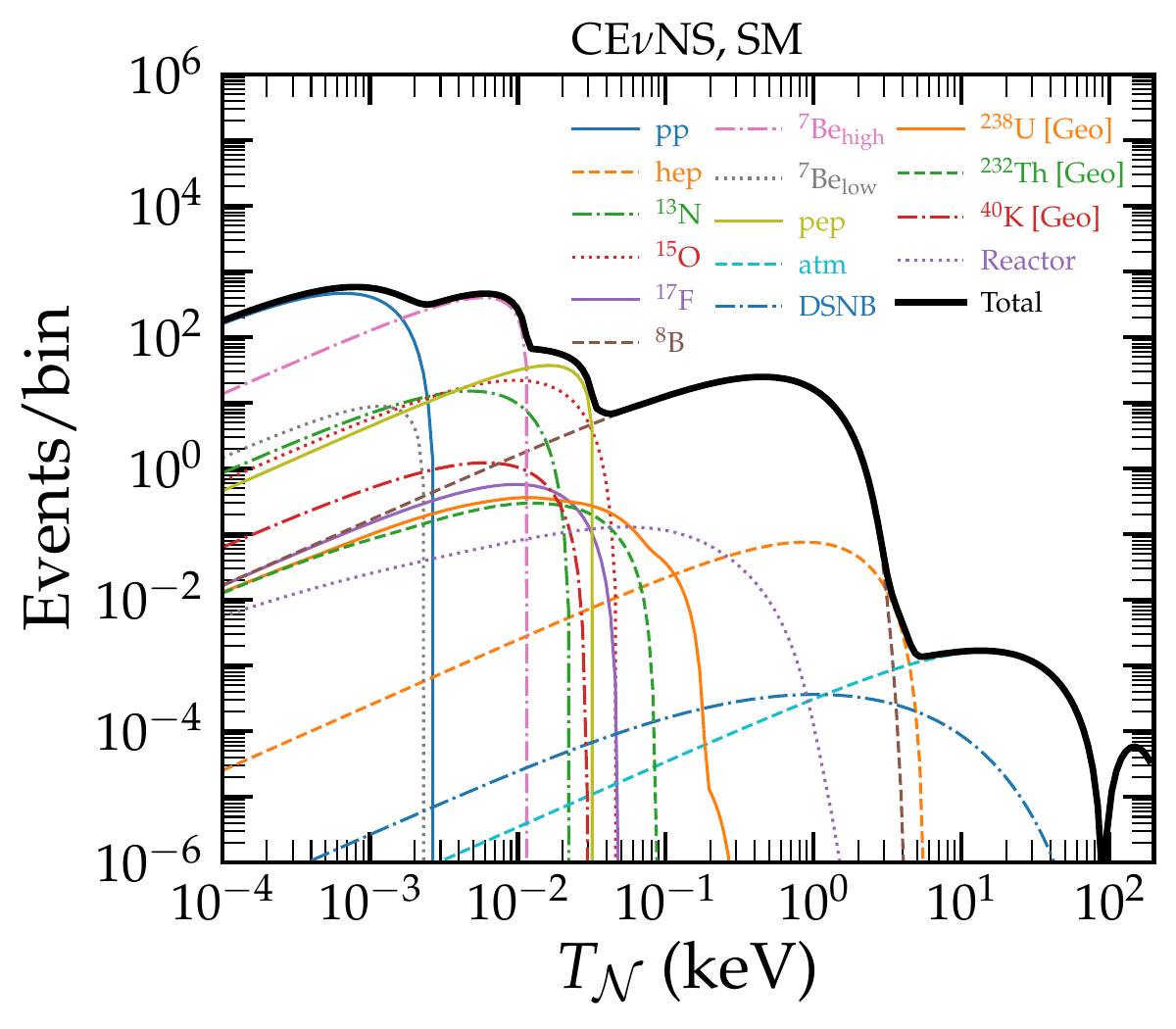}
  \includegraphics[width=0.48\textwidth]{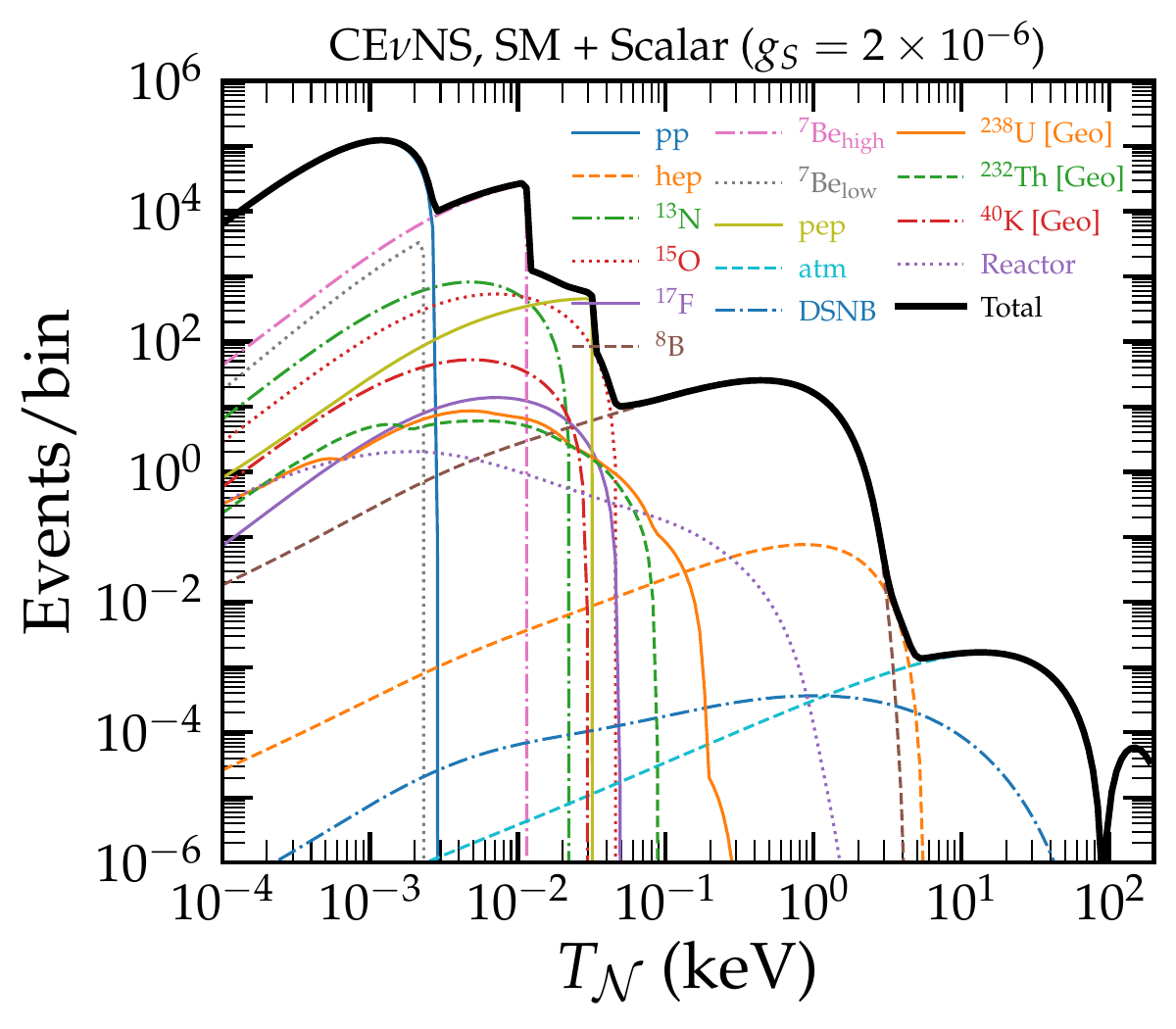}\\
    \includegraphics[width=0.48\textwidth]{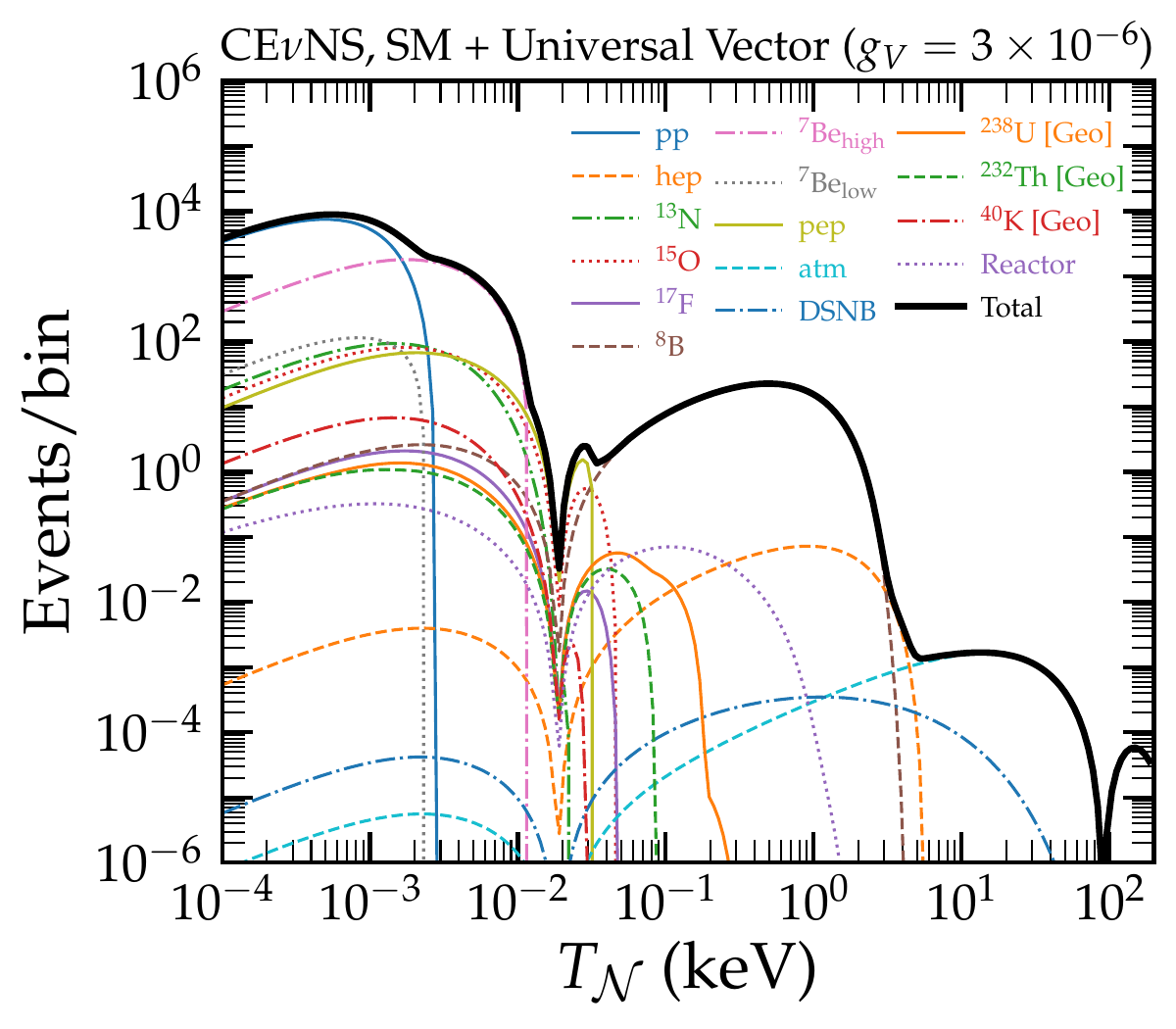}
  \includegraphics[width=0.48\textwidth]{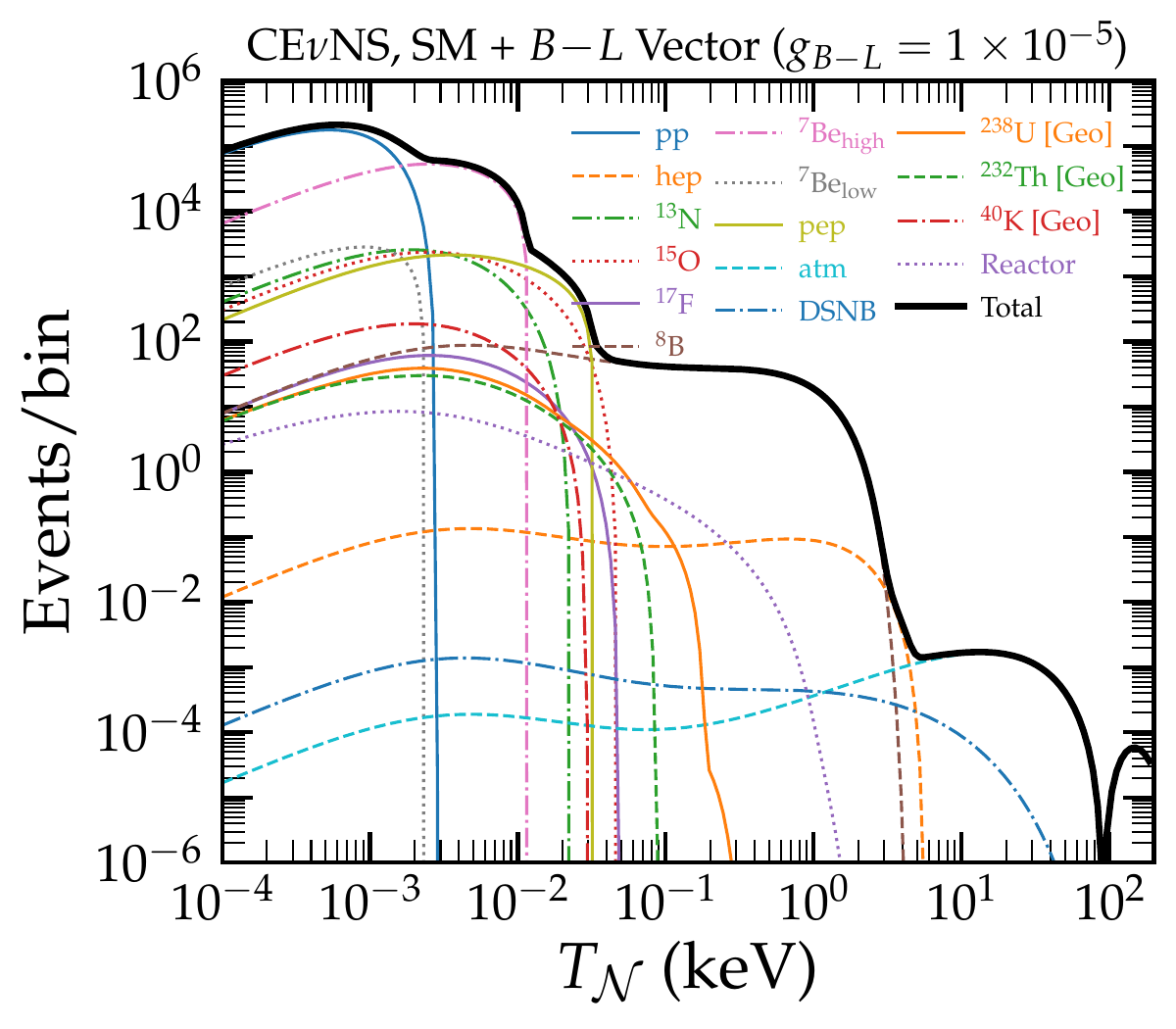}
  \caption{Expected \cevns\ event rates on a xenon target per recoil bin, assuming an exposure of $1~\mathrm{ton\times year}$ for the SM case and for several BSM interactions, assuming a mediator mass of $m_X = 1~\mathrm{MeV}$ and the couplings indicated in each panel title. The results are shown as individual recoil spectra for the various solar, DSNB, atmospheric, reactor, and geo-neutrino sources. The total predicted spectra are overlaid as thick black lines. For further details, see the main text.}
  \label{fig:CEvNS-rates}
\end{figure}

Fig.~\ref{fig:CEvNS-rates} shows the integrated \cevns~event rates  assuming SM interactions (upper left panel), or SM + a scalar mediator, with $m_S = 1$ MeV and $g_S = 2 \times 10^{-6}$ (upper right panel), SM + a vector mediator with universal couplings and $m_V = 1$ MeV and $g_V = 3 \times 10^{-6}$ (bottom left panel), and SM + a vector mediator with $B\!-\!L$ gauge symmetry, with $m_{B\!-\!L} = 1$ MeV and $g_{B\!-\!L} = 1 \times 10^{-5}$ (bottom right panel). In each panel, the different colored lines correspond to all neutrino fluxes possibly contributing to a signal in a Xe detector: solar, atmospheric, DSNB and geo-neutrinos. The thick black line indicates the total event rate. We can see that, compared to the SM case, in the presence of a new scalar or vector interaction, the rates are in general increased. The only exception is in the presence of a light vector mediator with universal couplings to neutrinos and quarks, which can also give rise to a destructive interference with the SM contribution, at $T_\mathcal{N} \sim 20~\mathrm{eV} $ for the chosen mediator coupling/mass benchmark values. Note that this value is  lower than the recoil-energy threshold of current DM direct detection experiments such as XENONnT. In full generality, the location of the dip can be identified as~\cite{AtzoriCorona:2022moj}

\begin{equation}
    T_\mathcal{N} = \frac{1}{2 m_\mathcal{N}} \left(\frac{3 \sqrt{2} (Z+N) g_V^2}{G_F \left[N - Z (1- 4 \sin^2 \theta_W )\right]} - m_V^2 \right) \, ,
    \label{eq:SM_cancellation}
\end{equation}
which can simplify further to $T_\mathcal{N}\approx  \frac{1}{2 m_\mathcal{N}} \left(\frac{3 \sqrt{2} (Z+N) g_V^2}{G_F N} - m_V^2 \right)$.
For a xenon target and BSM parameters $g_V=3 \times 10^{-6}$ and $m_V=1~\mathrm{MeV}$, it corresponds to the value $T_\mathcal{N} \approx 20~\mathrm{eV}$ mentioned above.
Moreover, the other dip visible at $T_\mathcal{N} \sim  90~\mathrm{keV} $ arises from the form factor suppression in the atmospheric neutrino flux.

\begin{figure}[htb!]
  \centering
  \includegraphics[width=0.48\textwidth]{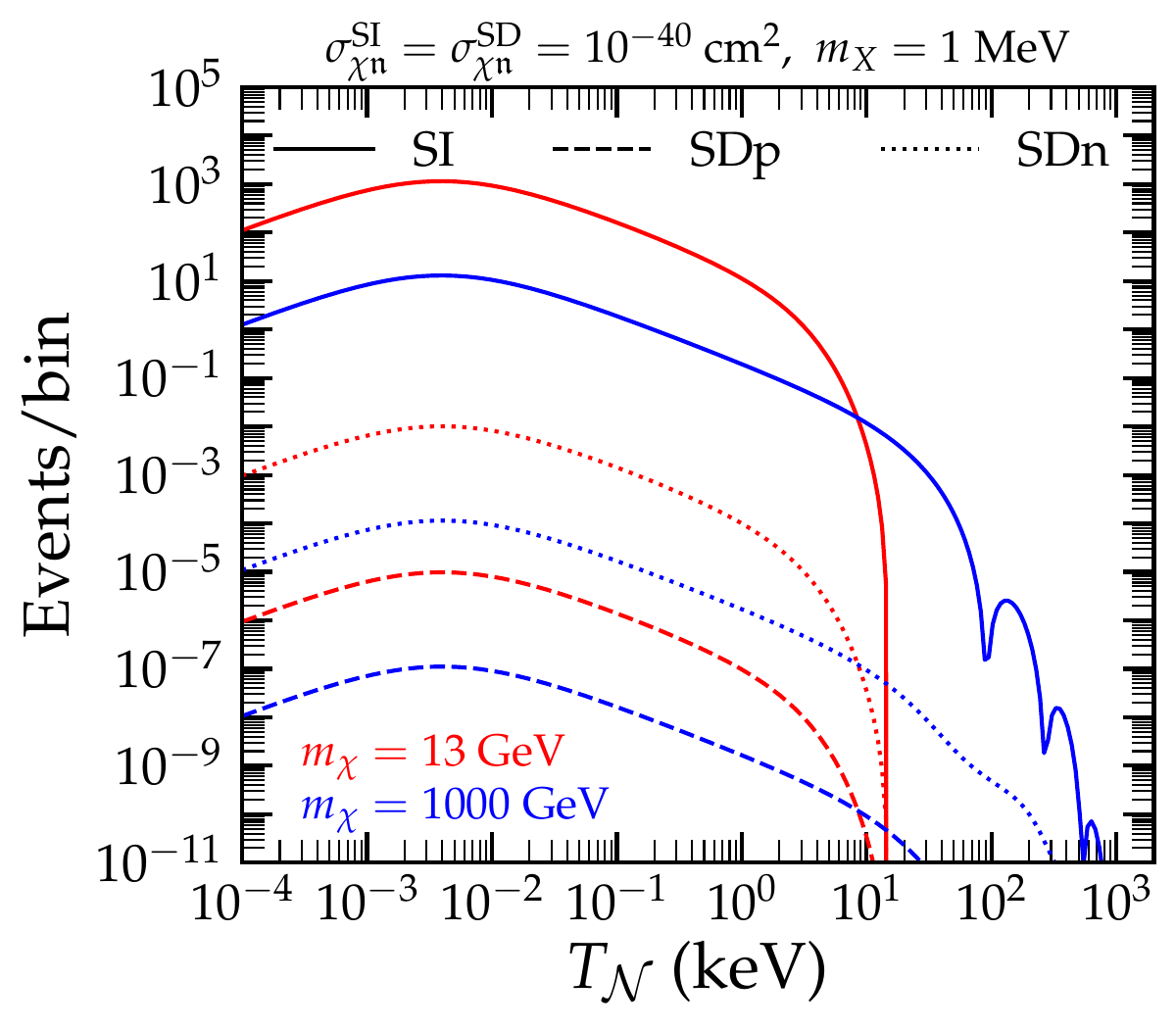}
  \includegraphics[width=0.48\textwidth]{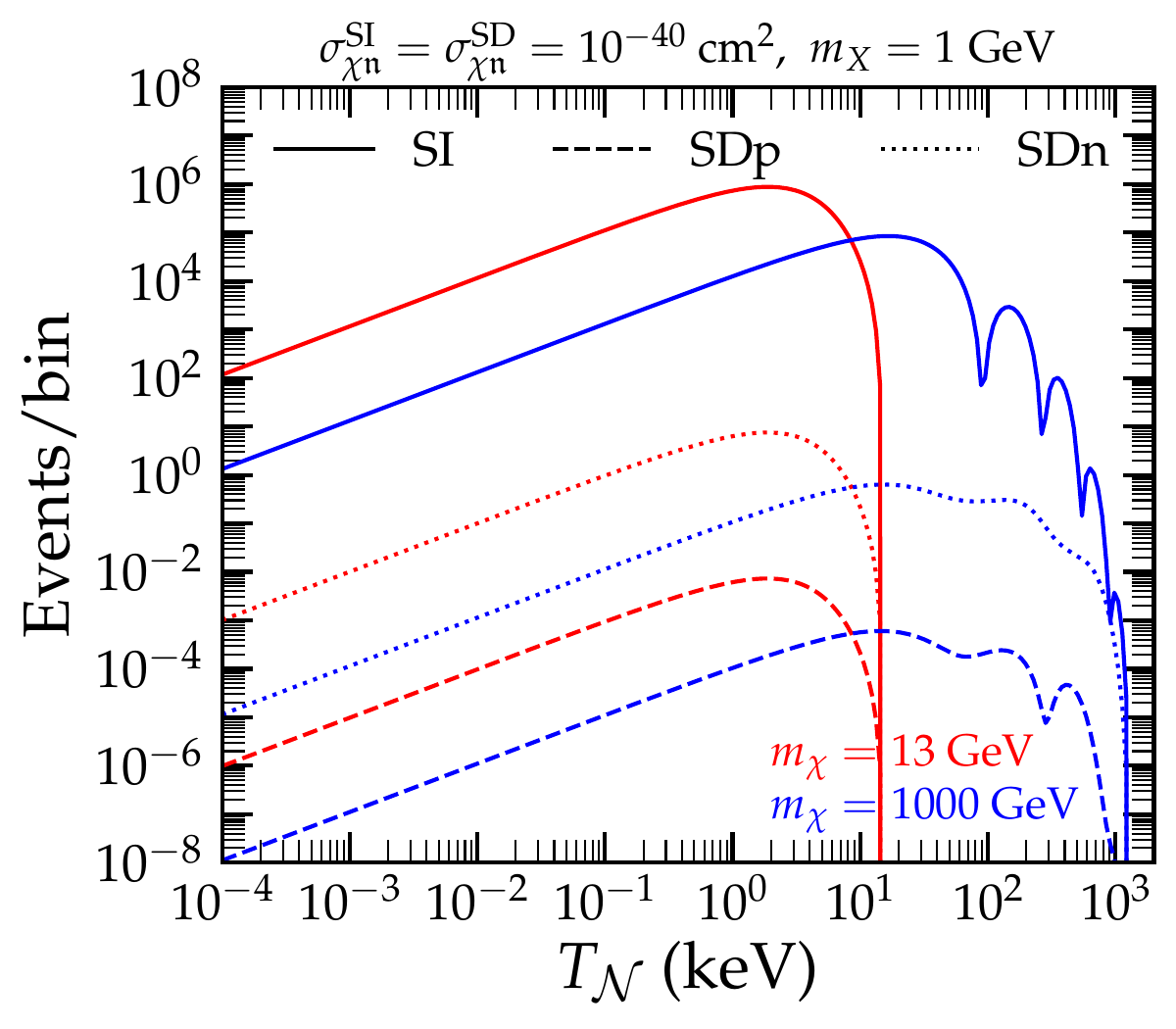}
  \caption{Expected SI and SD DM–nucleus event rates on a xenon target per recoil bin, assuming an exposure of $1~\mathrm{ton\times year}$ for two DM masses, $m_\chi = 13$ and $1000~\mathrm{GeV}$, and a representative DM–nucleon cross section of $\sigma_{\chi\mathfrak{n}}^{\mathrm{SI,(SD)}} = 10^{-40}~\mathrm{cm^2}$. The spectra are shown for a light mediator with mass $m_X = 1~\mathrm{MeV}$ ({\bf left panel}) and for a heavy mediator with mass $m_X = 1~\mathrm{GeV}$ ({\bf right panel}). For additional details, see the main text. }
  \label{fig:DM-rates}
\end{figure}

For comparison, Fig.~\ref{fig:DM-rates} illustrates the expected binned DM–nucleus event rates in scenario (i). The results are shown for the SI interactions (solid lines)  as well as for the two SD channels (proton-coupled, dashed; neutron-coupled, dotted), assuming both light and heavy mediators with masses $m_X = 1~\mathrm{MeV}$ and $m_X = 1~\mathrm{GeV}$, respectively, and a typical DM–nucleon cross section of $\sigma_{\chi\mathfrak{n}}^{\mathrm{SI,(SD)}} = 10^{-40}~\mathrm{cm^2}$. The event spectra are displayed for two benchmark DM masses, $m_\chi = 13$~GeV (red) and $1000$~GeV (blue).
One observes that for $m_\chi = 13$~GeV the predicted spectra are enhanced relative to the $m_\chi = 1000$~GeV case, which explains the stronger exclusion contours shown in Fig.~\ref{fig:BSM_Param_Space} as well as the XENONnT exclusion region in Figs.~\ref{fig:fog_SM}–\ref{fig:fog_caseii}. Additionally, the endpoints of the DM–nucleus spectra depend strongly on the DM mass, whereas for a fixed mass the SI and SD channels share the same endpoint, as dictated by kinematics.

Focusing on the SI spectra, significant suppression from the nuclear form factor appears as a series of dips starting from $T_\mathcal{N} \sim 90$~keV only for $m_\chi = 1000$~GeV. For $m_\chi = 13$~GeV, the kinematically available momentum transfer is too small for the form-factor suppression to become relevant. In the SD case, a mild suppression also emerges at $T_\mathcal{N} \sim 60$–$70$~keV, driven by the nuclear spin-structure function.

\bibliographystyle{utphys}
\bibliography{bibliography}

\end{document}